\documentclass[%
 aip,
 rsi,
amsmath,amssymb,
reprint,
floatfix
]{revtex4-1}

\usepackage{amsmath}
\usepackage{cases}
\usepackage{graphicx}
\usepackage{mathtools} % celtic 
\usepackage[T1]{fontenc} % celtic
\usepackage[utf8]{inputenc} % celtic
\usepackage{amsfonts} %celtic
\usepackage[toc,page]{appendix}

\usepackage{color}  %debug debug debug 

\begin{document}

%\preprint{}   

%\title{Nilpotent invariance of  CC energies: approximate inclusion of higher-rank excitations through nilpotent flow approach.}
%
% Coupled cluster formulations based on the nilpotent excitation sub-algebras
% Coupled cluster 
%

\title{Sub-system quantum dynamics using coupled cluster downfolding techniques}


\author{Karol Kowalski} 
\email{karol.kowalski@pnnl.gov}   
\author{Nicholas P. Bauman}          
%\thanks{}
\affiliation{Physical Sciences Division, Pacific Northwest National Laboratory, Richland, WA 99354, USA}


\date{\today}

\begin{abstract}
In this paper, we discuss extending the sub-system embedding sub-algebra coupled cluster (SESCC) formalism and the double unitary coupled cluster (DUCC) Ansatz to the time domain.
%revision_kk
%{\color{blue} 
An important part of the analysis is associated with proving the exactness of the DUCC Ansatz based on the general many-body form of anti-Hermitian cluster operators defining external and internal excitations.
%} 
Using these formalisms, it is possible to calculate the energy of the entire system as an eigenvalue of downfolded/effective 
Hamiltonian in the active space, that is identifiable with the sub-system of the composite system. It can also be shown  that downfolded Hamiltonians integrate out Fermionic degrees of freedom that do not correspond to the physics encapsulated by the active space.  
In this paper, we extend these results to the time-dependent Schr\"odinger equation, showing that 
a similar construct is possible to partition a system into a sub-system that varies slowly in time and a
remaining sub-system that corresponds to fast oscillations. This time-dependent formalism allows 
coupled cluster quantum dynamics to be extended to larger systems and for the formulation of novel quantum
algorithms based on the quantum Lanczos approach, which has recently been considered in the literature.

%GRA
\end{abstract}

%\keywords{}

\maketitle


\section{Introduction}
\label{section1}
The coupled cluster (CC) theory \cite{coester58_421,coester60_477,cizek66_4256,paldus72_50,purvis82_1910,paldus1999critical,
bartlett07_291} 
has evolved into one of the most accurate many-body formulations to 
describe correlated behavior of chemical \cite{bartlett07_291}  and nuclear systems.\cite{hagen2019doubly} Over the last few decades,
applications of the CC formalism in quantum chemistry  have grown enormously,  embracing molecular structure
optimization, the description of  chemical reactivity,
simulations of spectroscopic properties,
and computational models of  strongly correlated systems.
A great deal of effort has been exerted towards developing  
hierarchical families of approximations which provide an increasing level of accuracy by including high-rank collective
phenomena in cluster operator(s).\cite{bartlett07_291}
Significant advances  in describing properties, quasi-degenerate and excited electronic states  were possible  thanks to extensions of CC formalism 
to linear-response theory,\cite{monkhorst77_421,jorgensen90_3333} 
equation-of-motion CC formulations,
\cite{bartlett89_57,bartlett93_414,stanton1993equation,xpiecuch}  
and 
multi-reference CC methods.\cite{mukherjee1977applications,pal1988molecular,sinha1989eigenvalue,kaldor1991fock,meissner1998fock,musial2008multireference,jezmonk,meissner1,pylypov1,pylypov2,ims1,gms1,mahapatra1,mahapatra2,evangelista1,bwpittner1,lyakh12_182}
%GRA
 Significant progress has also been achieved in developing reduced scaling CC methods, mainly in applications
to ground- and excited-state problems.\cite{riplinger2013natural,riplinger2016sparse,peng2018state}  
The existence of hierarchical structures of approximations that allow one to reach the
exact, full configuration interaction (FCI), limit for a given basis set is an appealing feature 
of the CC formalism that drives the development of most formulations.
%GRA

Parallel to these advances, one could also witness  significant progress in developing 
explicitly time-dependent CC (TD-CC) formulations of the time-dependent Schr\"odinger equation (TDSE) 
\begin{equation}
i\hbar   \frac{\partial }{\partial t} |\Psi(t)\rangle = H |\Psi(t)\rangle\;,
\label{intro1}
\end{equation}
where
\begin{equation}
|\Psi(t)\rangle = e^{T(t)} |\Phi\rangle
\label{intro2}
\end{equation}
represents the time-dependent wave function 
with the time-dependent cluster operator $T(t)$.
This CC formulation has been explored in  CC linear-response theory \cite{monkhorst77_421,jorgensen90_3333,nascimento2016linear} for molecular systems, 
X-ray spectroscopy and Green's function theory,\cite{schonhammer1978time,nascimento2017simulation} nuclear physics,\cite{hoodbhoy1978time,hoodbhoy1979time,pigg2012time}  condensed matter physics,\cite{arponen1983variational}  and quantum dynamics of 
molecular systems in external fields.\cite{huber2011explicitly,kvaal2012ab,pedersen2019symplectic,kristiansen2020numerical,sato2018communication}
These studies have also initiated an intensive effort towards understanding many-aspects of the TD-CC formalism, including addressing fundamental problems such as 
the form of the action functional, form of the time-dependent molecular basis, various-rank approximations of the cluster operator, and numerical stability of 
time integration algorithms. One of the milestone achievements in developing time-dependent CC formalism was  Arponen's  action functional for the bi-variational coupled cluster formalism.
\cite{arponen1983variational}
In the last decade, this formalism was further extended by Kvaal  \cite{kvaal2012ab} by introducing the orbital adaptive time-dependent coupled cluster formalism 
and ensuing approximations.  
These developments made the TD-CC formalism a complementary approach to well established 
wave-function-based time-dependent    multi-configurational approaches,
\cite{meyer1990multi,beck2000multiconfiguration,nest2005multiconfiguration,miranda2011multiconfigurational,sato2013time,miyagi2014time,miyagi2014time2,peng2018simulating,liu2019time}
configuration interaction formulations,
\cite{sonk2011td,hochstuhl2012time,white2016computation,ulusoy2018role,lestrange2018time},
and density matrix renormalization group methods.
\cite{vidal2003efficient,white2004real,haegeman2016unifying,baiardi2019large}
%GRA

The sub-system embedding sub-algebra CC (SESCC) formalism 
\cite{safkk}
and its unitary variant based on the double unitary coupled cluster (DUCC) Ansatz
\cite{bauman2019downfolding}
enabled new  features of CC equations that are strictly related to the active space concept to be identified. The critical observation is related to the fact that energies of CC methods (such as CCSD,\cite{purvis82_1910} 
CCSDTQ,\cite{Kucharski1991,ccsdtq_nevin} 
etc.)  can be obtained, in contrast to the  standard CC energy expression, by  diagonalizing reduced-dimensionality effective (or downfolded) Hamiltonians  in the corresponding active space. 
%Both SESCC and DUCC methods provide   rigorous expressions defining these Hamiltonians. 
Additionally, downfolded Hamiltonians integrate out external Fermionic degrees of freedom (specifically, all cluster amplitudes 
that correspond to excitations outside of the active space). 
% revision_kk
%{\color{blue} 
For the DUCC case, this feature has been derived assuming the exactness of the double unitary CC Ansatz.
%}
%
Besides these fundamental properties, the SESCC formalism naturally introduces the concept of seniority numbers discussed recently in the context of  configuration interaction methods  
\cite{seniority1} 
and CC formulations.\cite{seniority2,seniority3,seniority4} 
The so-called SESCC flow equations \cite{safkk} and DUCC formalisms   were used to define approximations to calculate ground- and excited-states energies as well as  spectral functions in a recent 
Green's function DUCC extension.\cite{bauman2020coupled} The DUCC Hamiltonians  have also been  intensively tested on the subject of quantum computing simulations with reduced-dimensionality Hamiltonians.\cite{bauman2019downfolding,bauman2019quantum}
The SESCC methods complement/extend the active space coupled cluster ideas introduced in Refs. \cite{active3,activerev}  
(see also Refs. \cite{ active1, active2}),
 which also utilize the decomposition of the cluster operator into internal and external parts
 (for a detailed discussion of similarities and differences between active-space CC methods and SESCC see Refs.\cite{safkk,bauman2019downfolding}).
%GRA

In this manuscript, we present the time evolution of the system using  SESCC and DUCC  wave function representations. 
% revision_kk
%{\color{blue}
We also complement the discussion, showing that the exact wave function can be represented in the form of double unitary exponential Ansatz with general-type anti-Hermitian many-body cluster operators representing internal and external excitations. This result corresponds to the {\it general property} of the exact wave function proven at the level of SESCC formalism.
%}
%
As in previous studies, where SESCC/DUCC methods decoupled Fermionic degrees of freedom corresponding to various energy or localization regimes, we discuss formulations, using SESCC  and DUCC approaches, which
decouple slow- and fast-varying components of the wave function. 
Additionally, the  DUCC formalism  provides a rigorous many-body characterization of the time-dependent action functional to describe the  dynamics of the entire system 
in time modes captured by the corresponding active space. 
This approach and corresponding approximations can not only reduce the cost of 
TD-CC simulations for larger molecular applications 
but can also be employed in the imaginary time evolution, which has recently been intensively studied in the context of quantum computing.\cite{motta2020determining,mcardle2019variational} 
The flexibility associated with the choice of the active space can also be  advantageous for the generalization of time-dependent SESCC/DUCC 
formulations (TD-SESCC and TD-DUCC, respectively) beyond slow-varying components of the wave functions. 
In analogy to TD-CC formulations discussed in Refs.\cite{huber2011explicitly,pigg2012time}, we also analyze the properties of TD-SESCC and TD-DUCC methods based on fixed (time-independent) orthogonal spin orbitals. 
%GRA

\section{Downfolded CC Hamiltonians for a stationary Schr\"odinger equation}
\label{section2}
In this section, we  overview elements of SESCC and DUCC methods necessary in the analysis of TD-SESCC and TD-DUCC formalisms. 
For this to happen, let us  start with the time-independent formalism and summarize basic concepts behind sub-system embedding sub-algebras and the
double unitary CC expansion. 
%GRA

%%% revision_kk
\subsection{ Stationary SESCC formalism}
%%% revision_kk
The single reference CC (SR-CC) Ansatz  is predicated on the assumption that there exists a single Slater determinant $|\Phi\rangle$ that
provides a  reasonable approximation of the correlated  electronic ground-state  wave function $|\Psi\rangle$ to justify its exponential CC parametrization 
\begin{equation}
|\Psi\rangle = e^T |\Phi\rangle  \;,
\label{ccexp}
\end{equation}
where $T$ is the so-called cluster operator, which in general can be expressed in terms of its
many-body components $T_k$
\begin{equation}
T=\sum_{k=1}^{m} T_k \;.
\label{clustop}
\end{equation}
In the exact wave function limit,
the excitation level $m$  is equal to the number of correlated electrons ($N$) while in the approximate
formulations $m\ll N$. Several  standard approximations fall into this category, i.e.,  CCSD
($m=2$),\cite{purvis82_1910} CCSDT ($m=3$),
\cite{ccsdt_noga,ccsdt_noga_err,scuseria_ccsdt},
CCSDTQ ($m=4$),\cite{Kucharski1991,ccsdtq_nevin} etc.
Using the language of second quantization, the $T_k$ components can be expressed as
\begin{equation}
T_k = \frac{1}{(k!)^2} \sum_{i_1,\ldots,i_k; a_1\ldots a_k} t^{i_1\ldots i_k}_{a_1\ldots a_k} E^{a_1\ldots a_k}_{i_1\ldots i_k} \;,
\label{xex}
\end{equation}
where indices $i_1,i_2,\ldots$ ($a_1,a_2,\ldots$) refer to occupied (unoccupied) spin orbitals in the reference function $|\Phi\rangle$.
The excitation operators $E^{a_1\ldots a_k}_{i_1\ldots i_k} $ are defined through strings of standard creation ($a_p^{\dagger}$) and annihilation ($a_p$)
operators
\begin{equation}
E^{a_1\ldots a_k}_{i_1\ldots i_k}  = a_{a_1}^{\dagger}\ldots a_{a_k}^{\dagger} a_{i_k}\ldots a_{i_1} \;,
\label{estring}
\end{equation}
where 
creation and annihilation operators satisfy the following anti-commutation rules 
\begin{equation}
[a_p,a_q]_+ =
[a_p^{\dagger},a_q^{\dagger}]_+ = 0    \;, \label{comm1}
\end{equation}
\begin{equation}
[a_p,a_q^{\dagger}]_+ = \delta_{pq} \;.\label{comm2}
\end{equation}
After substituting   Ansatz (\ref{ccexp}) into the Schr\"odinger equation one gets the  energy-dependent form of the
CC equations:
\begin{equation}
(P+Q) He^T|\Phi\rangle= E (P+Q) e^T |\Phi\rangle \;\;,
\label{schreq}
\end{equation}
where $P$ and $Q$ are projection operators onto the reference function ($P=|\Phi\rangle\langle\Phi|$) and onto excited configurations 
(with respect to $|\Phi\rangle$) generated by
the $T$ operator when acting onto the reference function,
\begin{equation}
Q=\sum_{k=1}^{m}\;\sum_{i_1<i_2<\ldots<i_k; a_1<a_2\ldots <a_k}
 |\Phi_{i_1\ldots i_k}^{a_1 \ldots a_k}\rangle\langle \Phi_{i_1\ldots i_k}^{a_1 \ldots a_k}|  \;,
 \label{qoper}
\end{equation}
where 
\begin{equation}
|\Phi_{i_1\ldots i_k}^{a_1 \ldots a_k}\rangle = 
E^{a_1\ldots a_k}_{i_1\ldots i_k} |\Phi\rangle \;.
\label{exsl}
\end{equation}
Diagrammatic analysis \cite{paldus07} leads to an equivalent ({\it at the solution}), energy-independent form of the CC equations for cluster amplitudes
\begin{equation}
Qe^{-T}He^T|\Phi\rangle = Q(He^T)_C|\Phi\rangle = 0 \;,
\label{conform}
\end{equation}
and energy expression 
\begin{equation}
E=\langle\Phi|e^{-T} H e^T |\Phi\rangle
=\langle\Phi|(H e^T)_C |\Phi\rangle\;,
\label{eneex}
\end{equation}
where $C$ designates a connected part of a given operator expression. In the forthcoming discussion, we refer to 
$e^{-T}He^T$ as a similarity transformed Hamiltonian $\bar{H}$.

The SESCC formalism hinges upon the notion of excitation sub-algebra of commutative algebra $\mathfrak{g}^{(N)}$ generated by   
$E^{a_l}_{i_l}=a_{a_l}^{\dagger} a_{i_l}$ operators in the particle-hole representation (i.e,
$E^{a_1}_{i_1}=b^{\dagger}_{a_1} b^{\dagger}_{i_1}$ where 
$b^{\dagger}_{a_1}$ and $b^{\dagger}_{i_1}$ are particle and hole creation operators).
For detailed discussion of many-body Lie algebras the reader is referred to Refs.\cite{fukutome1981group,paldus_sarma,paldus_jeziorski}.
The SESCC formalism utilizes an important class of sub-algebras   of 
$\mathfrak{g}^{(N)}$,  which contain all possible excitations
$E^{a_1\ldots a_m}_{i_1\ldots i_m}$ that excite electrons from a subset of active occupied orbitals (denoted as $R$)
to a subset of active virtual orbitals (denoted as $S$).
These sub-algebras will be designated as $\mathfrak{g}^{(N)}(R,S)$.
In the following discussion, we will use $R$ 
%($R_i$, $i=1,2,\ldots,x$)  
and $S$ 
%($S_i$, $i=1,2,\ldots,y$) 
notation for subsets of occupied and virtual active orbitals $\lbrace R_i, \; i=1,\ldots,x \rbrace$ and 
$\lbrace S_i, \; i=1,\ldots,y \rbrace$, respectively (sometimes it is convenient to use alternative notation
$\mathfrak{g}^{(N)}(x_R,y_S)$ where numbers of active orbitals in $R$ and $S$ orbital  sets, $x$ and $y$, respectively,  are explicitly called out). 
As discussed in Ref.\cite{safkk} configurations  generated by elements of $\mathfrak{g}^{(N)}(x_R,y_S)$  along with the reference function 
span the complete active space (CAS) referenced to as the CAS($R,S$).

Each sub-algebra  $\mathfrak{h}=\mathfrak{g}^{(N)}(x_R,y_S)$ induces partitioning of the cluster operator into internal 
($T_{\rm int}(\mathfrak{h}$) or $T_{\rm int}$ for short) part belonging to $\mathfrak{h}$ and external 
($T_{\rm ext}(\mathfrak{h}$) or $T_{\rm ext}$ for short) part not belonging to $\mathfrak{h}$, i.e., 
\begin{equation}
T=T_{\rm int}(\mathfrak{h})+T_{\rm ext}(\mathfrak{h}) \;.
\label{decni}
\end{equation}
In Ref.\cite{safkk} it was shown that if two  criteria are met:
%\begin{enumerate}
(1) the $|\Psi(\mathfrak{h})\rangle= e^{T_{\rm int}(\mathfrak{h})}|\Phi\rangle$ is characterized by the same symmetry properties as 
$|\Psi\rangle$ and $|\Phi\rangle$ vectors
(for example,  spin and spatial symmetries),
and (2) the $e^{T_{\rm int}(\mathfrak{h})}|\Phi\rangle$ Ansatz generates FCI  expansion for the sub-system
defined by the CAS corresponding to the $\mathfrak{h}$ sub-algebra,
%\end{enumerate}
then $\mathfrak{h}$ is called a sub-system embedding sub-algebra (SES) for cluster operator $T$.
For any SES $\mathfrak{h}$ we proved the equivalence of two representations of the CC equations at the solution,
standard
\begin{eqnarray}
\langle\Phi| \bar{H}|\Phi\rangle &=& E \;,\label{rr00} \\
Q_{\rm int} \bar{H}|\Phi\rangle &=& 0  \;,\label{rr12} \\
Q_{\rm ext} \bar{H}|\Phi\rangle &=& 0 \;, \label{rr34}
\end{eqnarray}
and hybrid
\begin{eqnarray}
(P+Q_{\rm int}) \bar{H}_{\rm ext} e^{T_{\rm int}}|\Phi\rangle &=& E (P+Q_{\rm int})  e^{T_{\rm int}}|\Phi\rangle \;, \label{pp12} \\
Q_{\rm ext} \bar{H} |\Phi\rangle &=& 0 \;, \label{pp34}
\end{eqnarray}
where 
\begin{equation}
\bar{H}_{\rm ext}=e^{-T_{\rm ext}} H e^{T_{\rm ext}} 
\label{heffdef}
\end{equation}
and the two projection operators
$Q_{\rm int}(\mathfrak{h})$  and $Q_{\rm ext}(\mathfrak{h})$  ($Q_{\rm int}$ and $Q_{\rm ext}$ for short) are spanned by all excited configurations
generated by acting with $T_{\rm int}(\mathfrak{h})$ and $T_{\rm ext}(\mathfrak{h})$ onto reference function $|\Phi\rangle$,
respectively. The $Q_{\rm int}$ and $Q_{\rm ext}$ projections operators satisfy the condition
\begin{equation}
Q= Q_{\rm int}+ Q_{\rm ext}\;.
\label{qdec}
\end{equation}
The above equivalence shows that the CC energy can be calculated by diagonalizing effective Hamiltonian $H^{\rm eff}$ defined as 
\begin{equation}
H^{\rm eff}=(P+Q_{\rm int}) \bar{H}_{\rm ext} (P+Q_{\rm int})\;
\label{heffses}
\end{equation}
in the complete active space  corresponding to any  SES of  CC formulation defined by cluster operator $T$.
One should also notice that: (1)  the non-CAS related CC wave function components (referred here as external degrees of freedom)
are integrated out and encapsulated in the form of $H^{\rm eff}$, and (2) the internal part of the wave function,  $e^{T_{\rm int}}|\Phi\rangle$ is fully determined 
by diagonalization of $H^{\rm eff}$ in the CAS.
Separation of external degrees of freedom in the effective Hamiltonians is a desired feature especially from the point of view of building a
reduced-dimensionality Hamiltonian for quantum computing (QC). However, a factor that impedes  the use in QC of the $H^{\rm eff}$ is its 
non-Hermitian character.

%%% revision_kk
\subsection{ Stationary DUCC formalisms}
%%% revision_kk
In order to assure the Hermitian character of the CC effective Hamiltonian that also provides a separation of  Fermionic degrees of freedom,  in 
Ref.\cite{bauman2019downfolding} we have introduced double unitary coupled cluster Ansatz
\begin{equation}
        |\Psi\rangle=e^{\sigma_{\rm ext}} e^{\sigma_{\rm int}}|\Phi\rangle \;,
\label{ducc1}
\end{equation}
%%% revision_kk
%{\color{blue}    %%% revision_kk
where we assumed the exactness of expansion (\ref{ducc1}) 
in standard UCC parametrizaton
when all possible excitations are included in the definition of anti-hermitian 
$\sigma_{\rm ext}$ and $\sigma_{\rm int}$ operators, i.e., 
\begin{eqnarray}
\sigma_{\rm int}^{\dagger} &=&  -\sigma_{\rm int} \;, \label{sintah} \\
\sigma_{\rm ext}^{\dagger} &=&  -\sigma_{\rm ext} \;. \label{sintah2}
\end{eqnarray} 
Although numerical simulations \cite{evangelista2019exact} may suggest that the standard UCC parametrization can reproduce the exact wave function for 
model system, the generalization of this result to arbitrary  systems and to the doubly unitary CC parametrization remains unknown.
In this paper, we would like to fill this gap and prove that indeed there exist general  many-body $\sigma_{\rm ext}$ and $\sigma_{\rm int}$ operators
that reproduce the exact wave function when acting onto the reference function $|\Phi\rangle$. For this purpose, we will resort to the 
disentangled unitary coupled cluster methods introduced by Evangelista, Chan, and Scuseria.
\cite{evangelista2019exact}
This formalism   provides a powerful tool in the analysis 
of expansions based on non-commutative operator algebras. 
The main idea is to reduce the exact wave function $|\Psi\rangle$ to the reference determinant $|\Phi\rangle$
by  applying a series (sweeps)  of unitary transformations of the type 
\begin{equation}
e^{ \gamma^{\mathcal{O}}_{\mathcal{V}} (E^{\mathcal{V}}_{\mathcal{O}} -E^{\mathcal{O}}_{\mathcal{V}})} \;,
\label{binaryu}
\end{equation}
(where $\mathcal{O}$ and $\mathcal{V}$ denote  ordered strings of occupied $\lbrace i_1< \ldots < i_n \rbrace$ and 
virtual $\lbrace a_1 < \ldots < a_n \rbrace$ spinorbital indices)
that consecutively remove corresponding $|\Phi_{\mathcal{O}}^{\mathcal{V}}\rangle$ from wave function expansion. 
A key component of this algorithm is the ordering of these operations in the way that they do not reintroduce determinants that 
have already been removed. The process starts with the lowest occupied spinorbital "1" and unitary transformation that suppress 
the family of determinants 
\begin{equation}
\displaystyle\mathop{\forall}_{a}
 |\Phi_1^a\rangle \rightarrow 
 \displaystyle\mathop{\forall}_{i,a,b}
 |\Phi_{1i}^{ab} \rangle \rightarrow  
 \displaystyle\mathop{\forall}_{i,j,a,b,c}
 |\Phi_{1ij}^{abc} \rangle 
\rightarrow  \ldots \;,
\label{sweep1}
\end{equation}
were $\lbrace 1 i  \ldots \rbrace$ and  $\lbrace a b c \ldots \rbrace$ are ordered strings of occupied and virtual spinorbital indices.
In the next step one performs analogous operations for occupied spinorbital "2"
\begin{equation}
\displaystyle\mathop{\forall}_{a}
|\Phi_2^a\rangle \rightarrow 
\displaystyle\mathop{\forall}_{i,a,b}
|\Phi_{2i}^{ab} \rangle \rightarrow  
 \displaystyle\mathop{\forall}_{i,j,a,b,c}
 |\Phi_{2ij}^{abc} \rangle 
\rightarrow  \ldots \;,
\label{sweep2}
\end{equation}
etc., till all occupied spinorbital are exhausted. 
The final result  of applying the sequence of the above mentioned operations to $|\Psi\rangle$ is  the reference function $|\Phi\rangle$. 

\begin{figure}
\includegraphics[width=0.20 \textwidth]{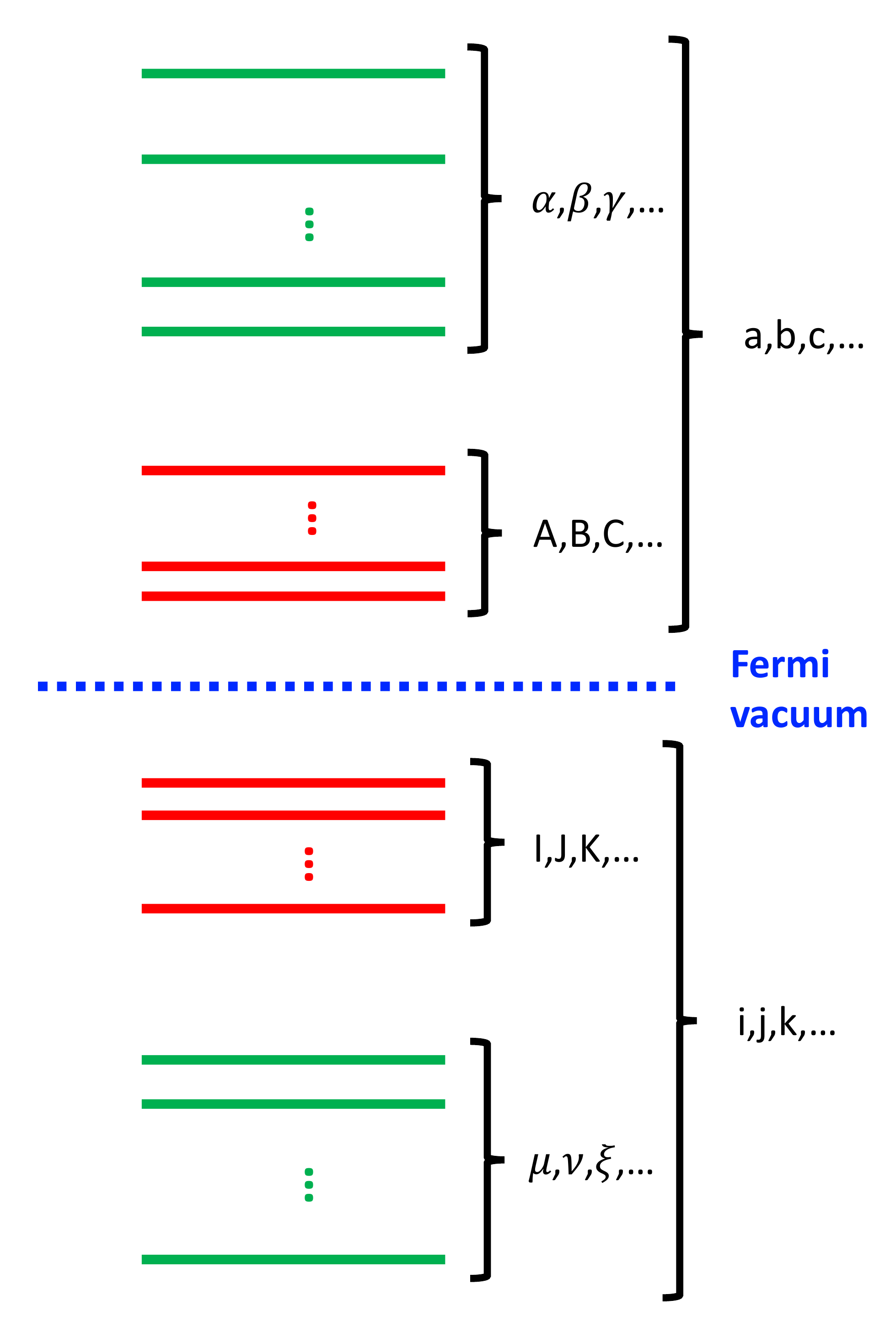}
\caption{ A schematic representation of the partitioning of  the spinorbital domain into occupied inactive ($\mu$,$\nu$,$\xi$,$\ldots$),
occupied active ($I$, $J$, $K$, $\ldots$), virtual active ($A$, $B$, $C$, $\ldots$), and virtual inactive ($\alpha$, $\beta$, $\gamma$,
$\ldots$). Generic occupied and virtual are denoted as $i$, $j$, $k$, $\ldots$, and $a$, $b$, $c$, $\ldots$, respectively.
}
\label{fig1}
\end{figure}

In order to prove the DUCC Ansatz for general many-body anti-Hermitian operators, let us modify the above procedure. For this purpose we will 
introduce partitioning of the occupied  and unoccupied spinorbitals into active and  inactive groups as shown in Fig.\ref{fig1} and we will  
enumerate them as  $\mu_1 < \dots < \mu_F < I_1 < \ldots < I_F < A_1 < \ldots <A_F < \alpha_1 < \ldots <\alpha_F$.
%\begin{widetext}
%\begin{center}
%\end{center}
%\end{widetext}
All Slater determinants spanning the external space can be partitioned into two disjoint sets of determinants 
$S_1=\lbrace |\Phi_{\mathcal{O}(\mu)}^{\mathcal{V}}\rangle   \rbrace_{\mu=\mu_1}^{\mu_F}$ and 
$S_2=\lbrace |\Phi_{\mathcal{O_{\rm act}}}^{\mathcal{V}(\alpha)}\rangle  \rbrace$, where $\mathcal{O}(\mu)$ and
$\mathcal{V}(\alpha)$ designate ordered string containing at least one inactive occupied  ($\mu$) and inactive virtual ($\alpha$) index, respectively, and 
$\mathcal{O}_{\rm act}$ represents strings of occupied active indices only. 

For the purpose of demonstrating the exactness of expansion (\ref{ducc1}), we will perform three classes of  sweeps:\\
%\begin{enumerate}
%\item 
{\bf Sweep 1.}
We perform rotations to suppress all external excitatons containing at least one occupied inactive index according to the flow:
\begin{eqnarray}
\displaystyle\mathop{\forall}_{a}
|\Phi_{\mu_1}^a\rangle &\rightarrow & 
\displaystyle\mathop{\forall}_{i,a,b}
|\Phi_{\mu_1i}^{ab} \rangle \rightarrow 
\displaystyle\mathop{\forall}_{i,j,a,b,c}
 |\Phi_{\mu_1ij}^{abc} \rangle   \rightarrow \ldots
\nonumber \\
\displaystyle\mathop{\forall}_{a}
|\Phi_{\mu_2}^a\rangle &\rightarrow & 
\displaystyle\mathop{\forall}_{i,a,b}
|\Phi_{\mu_2i}^{ab} \rangle \rightarrow 
\displaystyle\mathop{\forall}_{i,j,a,b,c}
 |\Phi_{\mu_2ij}^{abc} \rangle    \rightarrow \ldots \;,
\nonumber \\
\ldots && \ldots \nonumber \\
\displaystyle\mathop{\forall}_{a}
|\Phi_{\mu_F}^a\rangle &\rightarrow &
\displaystyle\mathop{\forall}_{i,a,b}
|\Phi_{\mu_Fi}^{ab} \rangle \rightarrow  
\displaystyle\mathop{\forall}_{i,j,a,b,c}
|\Phi_{\mu_Fij}^{abc} \rangle  \rightarrow \ldots  \;, 
\label{muspace}
\end{eqnarray}
where each step is given by basic unitary transformation (\ref{binaryu}).
In this way, through applying a finite product of elementary unitary transformations
\begin{equation}
\Omega_1=\prod_{\mu=\mu_F}^{\mu_1} \prod_{K_{\mu}} 
{\rm exp} \lbrace
\gamma^{\mathcal{O}(K_{\mu})}_{\mathcal{V}(K_{\mu})} 
(E^{\mathcal{V}(K_{\mu})}_{\mathcal{O}(K_{\mu})}
-E_{\mathcal{V}(K_{\mu})}^{\mathcal{O}(K_{\mu})})
\rbrace
\label{omega1}
\end{equation}
where all $\gamma^{\mathcal{O}(K_{\mu})}_{\mathcal{V}(K_{\mu})}$ are external excitations.  In other words, in $\Omega_1|\Psi\rangle$ all determinants belonging to $S_1$ have been eliminated. This step is identical with the first $\mu_F$ steps of the original algorithms if the ordering of occupied orbitals is the same.\\
%
%\item 
{\bf Sweep 2.}
In the second step, we perform elimination of Slater determinants in set $S_2$. This can be achieved by the following sequence of elimination steps involving basic 
unitary transformations:
\begin{eqnarray}
\displaystyle\mathop{\forall}_{\alpha}
|\Phi_{I_1}^{\alpha}\rangle &\rightarrow & 
\displaystyle\mathop{\forall}_{I',\alpha,b}
|\Phi_{I_1 I'}^{[\alpha b]} \rangle \rightarrow 
\displaystyle\mathop{\forall}_{I',I'',\alpha,b,c}
 |\Phi_{I_1 I' I''}^{[\alpha bc]} \rangle  \rightarrow \ldots
\nonumber \\
\displaystyle\mathop{\forall}_{\alpha}
|\Phi_{I_2}^{\alpha} \rangle &\rightarrow & 
\displaystyle\mathop{\forall}_{I',\alpha,b}
|\Phi_{I_2 I'}^{[\alpha b]} \rangle \rightarrow 
\displaystyle\mathop{\forall}_{I',I'',\alpha,b,c}
 |\Phi_{I_2 I' I''}^{[\alpha b c]} \rangle   \rightarrow \ldots
\nonumber \\
\ldots && \ldots \nonumber \\
\displaystyle\mathop{\forall}_{\alpha}
|\Phi_{I_F}^{\alpha} \rangle 
%&\rightarrow &
%\displaystyle\mathop{\forall}_{i,a,b}
%|\Phi_{\mu_Fi}^{ab} \rangle \rightarrow  
%\displaystyle\mathop{\forall}_{i,j,a,b,c}
%|\Phi_{\mu_Fij}^{abc} \rangle \rightarrow \ldots \;, 
\label{muspace}
\end{eqnarray}
where $I_k,I',I'', \ldots$ corresponds to the ordered strings 
$\lbrace I_k I' I''  \ldots \rbrace$, and $[\alpha b c \ldots]$ corresponds to 
%the summation over 
ordered strings containing elements $\alpha,b,c,\ldots$ (for example, if $\alpha<b$ then 
$[\alpha b]$ corresponds to $\lbrace \alpha b \rbrace $ otherwise to $\lbrace b \alpha \rbrace $).
The last step in (\ref{muspace}) corresponds to zeroing coefficients only  by single $|\Phi_{I_F}^a\rangle$ - 
since $I_F$ is the highest index of active occupied spinorbitals all other configurations
(for example $|\Phi_{I_F I'}^{[\alpha b]}\rangle$) were eliminated in earlier steps ($|\Phi_{I' I_F}^{[\alpha b]}\rangle$).
Sweep 2 corresponds to the following product of elementary unitary operations
\begin{equation}
\Omega_2=\prod_{I=I_F}^{I_1} \prod_{K_I} 
{\rm exp} 
\lbrace
\gamma^{O_{\rm act}(K_I)}_{V_{\alpha}(K_I)}
(
E_{O_{\rm act}(K_I)}^{V_{\alpha}(K_I)}
-E^{O_{\rm act}(K_I)}_{V_{\alpha}(K_I)}
)
\rbrace \;,
\label{omega2}
\end{equation}
where string $V_{\alpha}(K_{I})$ must contain at least one inactive virtual orbital $\alpha$.
As in Sweep 1, all amplitudes $\gamma^{O_{\rm act}(K_I)}_{V_{\alpha}(K_I)}$ are of the external type. 
After the Sweep 2 only surviving Slater determinants in $\Omega_2\Omega_1|\Psi\rangle$ belong to the active space, i.e.,
\begin{equation}
\Omega_2 \Omega_1 |\Psi\rangle = |\Psi_{\rm act}\rangle = (P+Q_{\rm int}) |\Psi_{\rm act}\rangle \;.
\end{equation} 
{\it This reduction of $|\Psi\rangle$ has been  achieved with the external excitations only. }
  \\
{\bf Sweep 3.} Repeating the reduction procedure for $|\Psi_{\rm act}\rangle$ in active spinorbital space we can write 
\begin{equation}
\Omega_3 |\Psi_{\rm act}\rangle = |\Phi\rangle \;,
\label{ract}
\end{equation}
 where $\Omega_3$ {\it involves rotations expressed in terms of internal excitations only}
\begin{equation}
\Omega_3 = \prod_{I=I_F}^{I_1} \prod_{K_I} 
{\rm exp} 
\lbrace
\gamma^{O_{\rm act}(K_I)}_{V_{\rm act}(K_I)}
(
E_{O_{\rm act}(K_I)}^{V_{\rm act}(K_I)}
-E^{O_{\rm act}(K_I)}_{V_{\rm act}(K_I)}
)
\rbrace \;,
\label{omega3}
\end{equation}
Summarizing, Sweeps 1-3 we can write 
\begin{equation}
|\Psi\rangle = \Omega_1^{-1} \Omega_2^{-1} \Omega_3^{-1} |\Phi\rangle \;,
\end{equation}
where $\Omega_1$ and $\Omega_2$ depends on the external excitations while $\Omega_3$ involves internal excitations only
(i.e., $\Omega_1=\Omega_1(\gamma_{\rm ext})$, $\Omega_2=\Omega_2(\gamma_{\rm ext})$, $\Omega_3=\Omega_3(\gamma_{\rm int})$). 
Applying multiple times Baker-Campbell-Hausdorff (BCH) formula 
\begin{equation}
e^X e^Y = e^{X+Y+\frac{1}{2}[X,Y]+\frac{1}{12}[X,[X,Y]]-\frac{1}{12}[Y,[X,Y]]+\ldots }
\label{bch}
\end{equation}
to $\Omega_3(\gamma_{\rm int})^{-1}$  and to the
product of $\Omega_1(\gamma_{\rm ext})^{-1} \Omega_2(\gamma_{\rm ext})^{-1}$ one obtains 
\begin{eqnarray}
\Omega_1(\gamma_{\rm ext})^{-1} \Omega_2(\gamma_{\rm ext})^{-1} &=& e^{-\gamma_{\rm ext}+C_{12}(\gamma_{\rm ext})} \label{sextp} \\
\Omega_3(\gamma_{\rm int})^{-1} &=& e^{\rm -\gamma_{\rm int} + C_3(\gamma_{\rm int} )} \label{sintp}
\end{eqnarray}
where $\gamma_{\rm ext}$ is a sum of external-type terms 
$\gamma^{\mathcal{O}}_{\mathcal{V}} (E^{\mathcal{V}}_{\mathcal{O}} -E^{\mathcal{O}}_{\mathcal{V}})$ defined in Sweeps 1 and 2 (see 
Eqs.(\ref{omega1}) and (\ref{omega2})) and 
$\gamma_{\rm int}$ is a sum of internal-type terms defined in Sweep 3 (Eq.\ref{omega3}). The $C_{12}(\gamma_{\rm ext})$ and $C_3(\gamma_{\rm int})$  are multiply-nested commutator expressions stemming from the multiple use of BCH expansion. 
Since $\gamma_{\rm ext}$ and $\gamma_{\rm int}$ are anti-Hermitian and the commutator of two anti-Hermitian operators is also anti-Hermitian. 
Therefore, we can represent the exact wave function in the form of product of the two unitary CC expansions involving external and internal degrees of freedom 
\begin{equation}
    |\Psi\rangle = e^{\sigma_{\rm ext}} e^{\sigma_{\rm int}} |\Phi\rangle
\label{duccp}
\end{equation}
where general-type anti-Hermitian external and internal operators $\sigma_{\rm ext}$ and $\sigma_{\rm int}$ are given by the expressions
\begin{eqnarray}
\sigma_{\rm ext} &=& -\gamma_{\rm ext}+C_{12}(\gamma_{\rm ext}) \;,
\label{extpar}\\
\sigma_{\rm int} &=& -\gamma_{\rm int}+C_{3}(\gamma_{\rm int}) \;.
\label{intpar}
\end{eqnarray}
Our following analysis will rely on the exactness of this expansion. If we change the parametrization
\begin{eqnarray}
\gamma_{\rm ext}&=& -(T_{\rm ext}-T_{\rm ext}^{\dagger}) \;,
\label{gtote}  \\
\gamma_{\rm int}&=& -(T_{\rm int}-T_{\rm int}^{\dagger}) \;,
\label{gtoti} 
\end{eqnarray}
where $T_{\rm ext}$ and $T_{\rm int}$ operators are characterized by the same  many-body structure as in the SESCC case (\ref{decni}), then 
\begin{eqnarray}
\sigma_{\rm ext} &=& T_{\rm ext}-T_{\rm ext}^{\dagger} +C_{12}(T_{\rm ext},T_{\rm ext}^{\dagger}) \;,
\label{extpar}\\
\sigma_{\rm int} &=& T_{\rm int}-T_{\rm int}^{\dagger}+C_{3}(T_{\rm int},T_{\rm int}^{\dagger}) \;.
\label{intpar}
\end{eqnarray}
Using the same arguments as in Ref.\cite{bauman2019downfolding} one can easily show that the exact DUCC expansion (\ref{duccp})
allows effective Hamiltonians to be constructed in a similar way as in single reference SESCC.
It can be proven that both the exact energy and the $e^{\sigma_{\rm int}}|\Phi\rangle$ FCI CAS state can be obtained by diagonalizing the DUCC effective Hamiltonian
in the complete active space
\begin{equation}
        H^{\rm eff} e^{\sigma_{\rm int}} |\Phi\rangle = E e^{\sigma_{\rm int}}|\Phi\rangle,
\label{duccstep2}
\end{equation}
where
\begin{equation}
        H^{\rm eff} = (P+Q_{\rm int}) \bar{H}_{\rm ext} (P+Q_{\rm int})
\label{equivducc}
\end{equation}
and 
\begin{equation}
        \bar{H}_{\rm ext} =e^{-\sigma_{\rm ext}}H e^{\sigma_{\rm ext}}.
\label{duccexth}
\end{equation}
%In analogy to the SESCC formalism, the energy and the CAS wave functions are obtained by diagonalizing DUCC effective Hamiltonian 
%in the complete active space. 
In the construction of the  DUCC effective Hamiltonian, only the external cluster operator 
($\sigma_{\rm ext}$) is used. 
In further analysis, for both SESCC and DUCC formalisms  we will  use the same notation for  the $\bar{H}_{\rm ext}$ and 
$H^{\rm eff}$ operators, and their form will follow from the  context of the discussed  equations, depending on if $T_{\rm ext}$, $\sigma_{\rm ext}$, or their analogous time-dependent variants $T_{\rm ext}(t)$ and $\sigma_{\rm ext}(t)$ are utilized. 

%revision2_kk
%{\color{blue}
The above analysis is predicated on the assumption that the infinite summations defining $\sigma_{\rm int}$ and $\sigma_{\rm ext}$ operators  (given by Eqs.(\ref{extpar}) and (\ref{intpar})) are convergent. For strongly correlated regime, 
where for example, amplitudes defining $T_{\rm int}$ are large, one can expect possible problems associated with the  convergence of expansion (\ref{intpar}). 
However, there are classes of applications such as the Quantum Phase Estimaton (QPE) algorithm in quantum computing,\cite{luis1996optimum,cleve1998quantum,berry2007efficient,childs2010relationship,wecker2015progress,haner2016high,poulin2017fast} 
where $\sigma_{\rm int}$'s  explicit construction 
is not required. Instead, the QPE algorithm  utilizes information about  $H^{\rm eff}$, which depends only on the $\sigma_{\rm ext}$ operator. By the proper choice of the active space, which should be large enough to make numerical values of $T_{\rm ext}$ sufficiently small, we can assume that the expansion (\ref{extpar}) is convergent.
In the following sections, we will also discuss other formulations 
(especially in the context of imaginary time evolution) where $e^{\sigma_{\rm int}}|\Phi\rangle$ Ansatz in the active space can be formally replaced by other exact expansions in the same active space. 
%}

A legitimate approximation of $\sigma_{\rm ext}$ and $\sigma_{\rm int}$
in Eqs.(\ref{extpar}) and (\ref{intpar})is to retain  lowest-order terms only, i.e.,
\begin{eqnarray}
\sigma_{\rm int} &\simeq& T_{\rm int} - T_{\rm int}^{\dagger} \;, \label{sint} \\
\sigma_{\rm ext} &\simeq& T_{\rm ext} - T_{\rm ext}^{\dagger} \;, \label{sext}
\end{eqnarray}
which has been discussed in Ref.\cite{bauman2019downfolding}.
%}  %%% revision_kk
%
%
% in kk_revision - justify this 
%
%
%
%
%
%
%
%
%

\section{Time-dependent formulations employing downfolded Hamiltonians}
\label{section3}

To  derive properties of the time-dependent Schr\"odinger equations utilizing
SESCC and DUCC  representations of the time-dependent wave functions,
we will  (in analogy to Refs.\cite{schonhammer1978time,huber2011explicitly,pigg2012time} focus on the  simplest case where orbitals and the reference function $|\Phi\rangle$ 
are  time-independent, which can be expressed as 
\begin{equation}
\frac{\partial}{\partial t} a_p = \frac{\partial}{\partial t} a_q^{\dagger}=0 \;, \; \frac{\partial}{\partial t} |\Phi\rangle =0 \;.
\label{orbnt}
\end{equation}
The above assumptions indicate  that  the CAS and corresponding SES $\mathfrak{h}$ do not change in time.

\subsection{Time-dependent Schr\"odinger equation in the SESCC representation} 

In this subsection, we derive the time-dependent extension of the SESCC and DUCC wave function representations. 
First, we start from the time-dependent CC parametrization of the wave function:
\begin{equation}
|\Psi_{\rm CC}(t)\rangle = e^{T(t)} |\Phi\rangle \;,
\label{cct}
\end{equation}
where 
\begin{equation}
T(t)=\sum_{k=0}^{m} T_k(t) \;.
\label{clustop2}
\end{equation}
As in the stationary SESCC formulation, we will assume the decomposition of the time-dependent cluster operator $T(t)$ into 
internal ($T_{\rm int}(t)$) and external ($T_{\rm ext}(t)$) parts, i.e,
\begin{eqnarray}
|\Psi_{\rm CC}(t)\rangle &=&  e^{T_{\rm ext}(t)+T_{\rm int}(t)} |\Phi\rangle \label{step1} \\
&=& e^{T_{\rm ext}(t)} e^{T_{\rm int}(t)} |\Phi\rangle \;. \label{step2} 
\end{eqnarray}
%% initial conditions ---> define
%
For the sake of generality, we  also include a time-dependent scalar phase factor ($T_0(t)$) 
in the definition of the  $T_{\rm int}(t)$ operator.
As pointed out by Hoodbhoy and Negele in Refs.\cite{hoodbhoy1978time,hoodbhoy1979time}
this phase factor is  not needed when calculating 
physical observables.

Upon introducing expansion (\ref{step2}) into the time-dependent Schr\"odinger equation, one obtains 
\begin{equation}
i\hbar \frac{\partial }{\partial t} e^{T_{\rm ext}(t)} e^{T_{\rm int}(t)} |\Phi\rangle = H e^{T_{\rm ext}(t)} e^{T_{\rm int}(t)} |\Phi\rangle \;,
\label{tdepcc}
\end{equation}
which can be further transformed (after differentiating its left-hand side over $t$)
\begin{eqnarray}
i\hbar (\frac{\partial T_{\rm int}(t)}{\partial t} + \frac{\partial T_{\rm ext}(t))}{\partial t})  e^{T_{\rm ext}(t)} e^{T_{\rm int}(t)} |\Phi\rangle &=& \nonumber \\
&& \hspace*{-6.0cm} H e^{T_{\rm ext}(t)} e^{T_{\rm int}(t)} |\Phi\rangle \;,
\label{tdepcc2}
\end{eqnarray}
After pre-multiplying both sides by $e^{-T_{\rm ext}(t)}$,  we obtain a convenient form of TDSE
\begin{eqnarray}
i\hbar (\frac{\partial T_{\rm int}(t)}{\partial t} + \frac{\partial T_{\rm ext}(t)}{\partial t} ) e^{T_{\rm int}(t)} |\Phi\rangle  
 &=& \nonumber \\ 
&& \hspace{-3.0cm} \bar{H}_{\rm ext}(t) e^{T_{\rm int}(t)}|\Phi\rangle \;,
\label{intermed1}
\end{eqnarray}
where 
\begin{equation}
\bar{H}_{\rm ext}(t) = e^{-T_{\rm ext}(t)} H e^{T_{\rm ext}(t)} \;.
\label{hbart}
\end{equation}
Now,   we focus our attention on the projection of Eq.(\ref{intermed1}) onto time-independent subspace $(P+Q_{\rm int})$, which leads to the 
equations:
\begin{eqnarray} 
i\hbar  (P+Q_{\rm int}) (\frac{\partial T_{\rm int}(t) }{\partial t} + \frac{\partial T_{\rm ext}(t)}{\partial t}) e^{T_{\rm int}(t)} |\Phi\rangle && \nonumber \\
=
(P+Q_{\rm int}) \bar{H}_{\rm ext}(t)  e^{T_{\rm int}(t)} |\Phi\rangle \;.&&
\end{eqnarray}
Taking into account the fact that  $T_{\rm ext}(t)$ and $ \frac{\partial }{\partial t} T_{\rm ext}(t)$ produce "external" excitations (defined by strings of creation-annihilation operators containing at least one inactive spin-orbital index) we have 
\begin{equation}
(P+Q_{\rm int})  \frac{\partial T_{\rm ext}(t)}{\partial t}  e^{T_{\rm int}(t)} |\Phi\rangle = 0 
\end{equation}
and therefore the  time evolution of $e^{T_{\rm int}(t)} |\Phi\rangle$  corresponds to the  non-unitary evolution in 
time-independent  $(P+Q_{\rm int})$ space
\begin{equation}
i\hbar  (P+Q_{\rm int}) \frac{\partial T_{\rm int}(t)}{\partial t}  e^{T_{\rm int}(t)} |\Phi\rangle = 
(P+Q_{\rm int}) \bar{H}_{\rm ext}(t)  e^{T_{\rm int}(t)} |\Phi\rangle  
\label{cool1}
\end{equation}
or 
\begin{equation}
i\hbar \frac{\partial }{\partial t} e^{T_{\rm int}(t)} |\Phi\rangle = H^{\rm eff}(t)  e^{T_{\rm int}(t)} |\Phi\rangle \;,
\label{cool2}
\end{equation}
where we used the fact that 
\begin{equation}
(P+Q_{\rm int}) e^{T_{\rm int}(t)} |\Phi\rangle = e^{T_{\rm int}(t)} |\Phi\rangle
\label{lemma1}
\end{equation}
and 
\begin{equation}
H^{\rm eff}(t) = (P+Q_{\rm int}) \bar{H}_{\rm ext}(t) (P+Q_{\rm int}) \;.
\label{lemma2}
\end{equation}
We will refer to equation (\ref{cool2}) as  embedded sub-system  time evolution for the ket-state.
%{\color{blue} 
In the above equations, we avoided the explicit operator notation $i\hbar
\frac{\partial }{\partial t} T_{\rm int}(t)|\Phi\rangle=e^{-T_{\rm int}(t)} H^{\rm eff}(t) e^{T_{\rm int}(t)}|\Phi\rangle$ since the same notation is much harder to define in the DUCC case.
%}
% revision_kk some disucssion of ub-system needed here.

%%{\color{red} Discussion of explicit utilization of the Schr\"odinger equations as in Klammath paper, Gunnarson, Negele.
%%Insert discussion of the sub-system dynamic in this section}

\subsection{Time-dependent Schr\"odinger equation in the DUCC representation} 

The non-Hermitian character of $\bar{H}_{\rm ext}(t)$ may limit 
%revision_kk
%{\color{blue} 
applications of this formalism  in the area of quantum computing.
%}
In analogy to the 
DUCC formalism studied in Refs.\cite{bauman2019downfolding,bauman2019quantum,bauman2020coupled} let us represent normalized time-dependent wave function 
$|\Psi_{\rm DUCC}(t)\rangle$ in the following form 
\begin{equation}
|\Psi_{\rm DUCC}(t)\rangle = e^{\sigma_{\rm ext}(t)} e^{\sigma_{\rm int}(t)} |\Phi\rangle \;,
\label{ducct}
\end{equation}
% shows the intensive character of excited states
where $\sigma_{\rm int}(t)$ and $\sigma_{\rm ext}(t)$ are 
%revision_kk
%{\color{blue} 
general-type time-dependent anti-Hermitian operators in the sense of Eqs.(\ref{extpar}) and (\ref{intpar}), i.e.,
%}
\begin{eqnarray}
\sigma_{\rm int}(t)^{\dagger} &=& - \sigma_{\rm int}(t)\;,  \label{ucc1} \\
\sigma_{\rm ext}(t)^{\dagger} &=& - \sigma_{\rm ext}(t)\;.  \label{ucc2}
\end{eqnarray}
% revision_kk
%{\color{blue} 
We also assume that the phase factor $i\delta$ is  included in the definition of $\sigma_{\rm int}$.
%}
% revision_kk
Introducing (\ref{ducct}) into the TDSE leads to the equation
\begin{eqnarray}
i\hbar (\frac{\partial e^{\sigma_{\rm ext}(t)} }{\partial t}  e^{\sigma_{\rm int}(t)}+  
e^{\sigma_{\rm ext}(t)} \frac{\partial e^{\sigma_{\rm int}(t)}}{\partial t} ) |\Phi\rangle &=& \nonumber \\
&&\hspace*{-2.6cm} H e^{\sigma_{\rm ext}(t)}  e^{\sigma_{\rm int}(t)} |\Phi\rangle  \;. \label{sigmat}
\end{eqnarray}
% revision_kk
%{\color{blue}
By pre-multiplying both sides of the above equation by 
$e^{-\sigma_{\rm ext}(t)}$ and projecting onto 
$(P+Q_{\rm int})$ subspace one obtains
\begin{widetext}
\begin{equation}
i\hbar \frac{\partial e^{\sigma_{\rm int}(t)}}{\partial t}  |\Phi\rangle = 
H^{\rm eff}(\sigma_{\rm ext}(t),\frac{\partial \sigma_{\rm ext}(t)}{\partial t}) 
e^{\sigma_{\rm int}(t)}
|\Phi\rangle \;,
\label{heffrev1}
\end{equation}
where 
\begin{equation}
H^{\rm eff}(\sigma_{\rm ext}(t),\frac{\partial \sigma_{\rm ext}(t)}{\partial t})=
(P+Q_{\rm int})\lbrack 
\bar{H}_{\rm ext} 
-i \hbar e^{-\sigma_{\rm ext}(t)} 
\frac{\partial e^{\sigma_{\rm ext}(t)}}{\partial t}
\rbrack (P+Q_{\rm int}) 
\label{heffrev2}
\end{equation}
\end{widetext}
and 
\begin{equation}
\bar{H}_{\rm ext}(t)=e^{-\sigma_{\rm ext}(t)} H e^{\sigma_{\rm ext}(t)} \;.
\end{equation} 
%}
%In contrast to the standard  single reference formalism, where all $T$ operators are represented by commuting components and 
%calculations of time derivatives of exponential operators are pretty straightforward, the unitary case, where components of 
%$\sigma_{\rm ext}(t)$ and $\sigma_{\rm int}(t)$ are non-commuting operators,   poses a significant challenge.
%
% revision_kk
%{\color{blue}
To analyze the many-body structure of the Hermitian effective Hamiltonian (\ref{heffrev2})
%}
we will use the  following identity  for calculating  derivatives of exponential operators (see Refs.\cite{rossmann2006lie,hall2015lie} for more details)
\begin{eqnarray}
\frac{\partial }{\partial t} e^{X(t)} &=&  e^{X(t)} \sum_{k=0}^{\infty} \frac{(-1)^k}{(k+1)!} (ad_{X(t)})^k \frac{\partial X(t)}{\partial t}
\label{sned1} \\
&=& e^{X(t)} \sum_{k=0}^{\infty} \frac{(-1)^k}{(k+1)!} I_k(X(t),\frac{\partial X(t)}{\partial t})
\label{sned2} 
\end{eqnarray}
where the adjoint action $ad_X$ is defined as 
\begin{equation}
ad_A(B)=[A,B] \;
\label{adx}
\end{equation}
and $k$-commutator term $I_k(X(t),\frac{\partial X(t)}{\partial t})$ is given by the formula
\begin{equation}
I_k(X(t),\frac{\partial X(t)}{\partial t}) = \underbrace{[X(t),[X(t), \ldots  [X(t),[X(t),\frac{\partial X(t)}{\partial t}]]\ldots ]]}_{k \; {\rm times}}\;.
\label{ik}
\end{equation}
It is easy to notice that the first terms in the expansion (\ref{sned2}) are given by  the expressions 
\begin{eqnarray}
I_0(X(t),\frac{\partial X(t)}{\partial t})&=&\frac{\partial X(t)}{\partial t} \;, \label{i0} \\
I_1(X(t),\frac{\partial X(t)}{\partial t})&=&[X(t),\frac{\partial X(t)}{\partial t}] \;,
\label{i1}
\end{eqnarray}
and for $k\ge1$ recursive formula is satisfied
\begin{equation}
I_k=[X(t),I_{k-1}(t)] \;.
\label{ik2}
\end{equation}
Henceforth, for the simplicity of notation, we will use $I_k$ and 
$I_k(X(t),\frac{\partial X(t)}{\partial t})$ interchangeably. 

Formula (\ref{sned2}) will be used to evaluate the
\begin{equation}
\frac{\partial e^{\sigma_{\rm ext}(t)} }{\partial t}  
\label{bubu1}
\end{equation} 
term in Eq.(\ref{heffrev2}), which using formula (\ref{sned2}) can be re-written in the form: 
\begin{equation} 
e^{\sigma_{\rm ext}(t)} A(\sigma_{\rm ext}(t), \frac{\partial \sigma_{\rm ext}(t)}{\partial t}) 
\label{dera}
\end{equation} 
where
\begin{equation}
A(\sigma_{\rm ext}(t), \frac{\partial \sigma_{\rm ext}(t)}{\partial t}) = 
\sum_{k=0}^{\infty} \frac{(-1)^k}{(k+1)!} I_k(\sigma_{\rm ext}(t), \frac{\partial \sigma_{\rm ext}(t)}{\partial t}) \;.
\label{asigt}
\end{equation} 
Given the fact that both $\sigma_{\rm ext}(t)$ and $\frac{\partial \sigma_{\rm ext}(t)}{\partial t}$ operators are anti-Hermitian, 
it is easy to show the same is true for $I_k(\sigma_{\rm ext}(t), \frac{\partial \sigma_{\rm ext}(t)}{\partial t})$ and 
$A(\sigma_{\rm ext}(t), \frac{\partial \sigma_{\rm ext}(t)}{\partial t})$ operators, i.e.,
\begin{eqnarray}
I_k(\sigma_{\rm ext}(t), \frac{\partial \sigma_{\rm ext}(t)}{\partial t})^{\dagger} &=& 
-I_k(\sigma_{\rm ext}(t), \frac{\partial \sigma_{\rm ext}(t)}{\partial t}) \;, \label{ikdagg} \\
A(\sigma_{\rm ext}(t), \frac{\partial \sigma_{\rm ext}(t)}{\partial t})^{\dagger} &=& 
-A(\sigma_{\rm ext}(t), \frac{\partial \sigma_{\rm ext}(t)}{\partial t}) \;. \label{adagg} 
\end{eqnarray}
%By introducing Eq.(\ref{dera}) into Eq.(\ref{sigmat}),
%multiplying both sides by $e^{-\sigma_{\rm ext}(t)}$,
%and projecting resulting equations onto $(P+Q_{\rm int})$, one gets the equation 
%for time evolution of the sub-system $e^{\sigma_{\rm int}(t)}|\Phi\rangle$
%\begin{equation}
%i\hbar \frac{\partial}{\partial t} e^{\sigma_{\rm int}(t)}|\Phi\rangle =
%H^{\rm eff}(\sigma_{\rm ext}(t),\frac{\partial \sigma_{\rm ext}(t)}{\partial t}) \;\; e^{\sigma_{\rm int}(t)}|\Phi\rangle \;,
%\label{fsub}
%\end{equation}
Now, the  effective (or downfolded) Hamiltonian (\ref{heffrev2})
%$H^{\rm eff}(\sigma_{\rm ext}(t),\frac{\partial \sigma_{\rm ext}(t)}{\partial t})$  
is given be the expression
%\begin{eqnarray}
%H^{\rm eff}(\sigma_{\rm ext}(t),\frac{\partial \sigma_{\rm ext}(t)}{\partial t}) = &&\nonumber \\
%&&
%\hspace*{-4.0cm} (P+Q_{\rm int} )\lbrace  \bar{H}_{\rm ext}(t) %-i\hbar 
%A(\sigma_{\rm ext}(t), \frac{\partial \sigma_{\rm ext}(t)}{\partial t})  \rbrace
%(P+Q_{\rm int})  
%\label{gammae} 
%\end{eqnarray}
\begin{widetext}
\begin{equation}
H^{\rm eff}(\sigma_{\rm ext}(t),\frac{\partial \sigma_{\rm ext}(t)}{\partial t}) = 
(P+Q_{\rm int} )\lbrace  \bar{H}_{\rm ext}(t) -i\hbar 
A(\sigma_{\rm ext}(t), \frac{\partial \sigma_{\rm ext}(t)}{\partial t})  \rbrace
(P+Q_{\rm int})  \;.
\label{gammae} 
\end{equation}
\end{widetext}
%
%and
%\begin{equation}
%\bar{H}_{\rm ext}(t)=e^{-\sigma_{\rm ext}(t)} H e^{\sigma_{\rm ext}(t)} \;.
%\end{equation} 
In deriving the above equations,  we employed the fact that 
\begin{equation}
e^{\sigma_{\rm int}(t)}|\Phi\rangle = (P+Q_{\rm int}) e^{\sigma_{\rm int}(t)}|\Phi\rangle \;.
\label{pqint}
\end{equation}
%One should also notice that 
%$H^{\rm eff}(\sigma_{\rm ext}(t),\frac{\partial \sigma_{\rm ext}(t)}{\partial t})$ 
%is an Hermitian operator. 
%
%
%
%
%

\subsection{Common features of TD-SESCC and TD-DUCC formulations}
In  previous subsections, we showed that  when the orbital basis is time-independent then for  both TD-SESCC and TD-DUCC cases the dynamics of the sub-system at the level of ket-state 
(defined by appropriately chosen 
SES $\mathfrak{h}$ or equivalently active space)  can be described by the effective/downfolded Hamiltonians acting in the active space. 
While in the TD-SESCC case the dynamic is generated by a  non-Hermitian Hamiltonian, for the TD-DUCC formalism, the downfolded Hamiltonian is Hermitian and contains 
"external-velocity" dependent term (i.e., $\frac{\partial \sigma_{\rm ext}(t)}{\partial t}$-dependent term  - $A$ operator in Eq.(\ref{gammae})). 
In both cases 
%revision_kk
%{\color{blue} 
at the level of the ket-state evolution,
%}
one can observe a rigorous decomposition of external  Fermionic degrees of freedom ($T_{\rm ext}(t)$ and $\sigma_{\rm ext}(t)$ operators) from those defining 
sub-system time-dependent wave function ($T_{\rm int}(t)$ and $\sigma_{\rm int}(t)$) in the time-dependent downfolded Hamiltonian. 
By the appropriate choice of the active space, these approaches
can be used to separate the description of the  sub-system that slowly evolves  in time (representing low-energy modes)  from the components of the entire system that correspond to fast-varying components (representing high-energy modes) of the entire
system. As suggested by earlier analysis (see Ref.\cite{bauman2019downfolding}) parameters corresponding to fast-varying parts  can be effectively described by 
perturbation techniques. Simulations  of these sub-systems can be performed  employing explicit time-propagation  techniques described in Refs.\cite{schonhammer1978time,hoodbhoy1978time,hoodbhoy1979time,huber2011explicitly}.

%revision_kk
%{\color{blue}
Although  it has  already been  discussed in the literature (for example, see Ref.\cite{kvaal2012ab}) that to calculate expectation values of  physical observables in the CC method,  additional state parameters need to be introduced (in the case of the standard CC theory these parameters correspond to  cluster operator $T$ and the so-called $\Lambda$ operator, where $\langle\Phi|(1+\Lambda)$ is the   left eigenvector of the $\bar{H}$ operator),  the above decomposition of the ket-variant of 
TDSE plays an important role for newly introduced equation-of-motion CC cumulant Green's function theory \cite{rehr_cumulant} especially in the context of  calculating spectral functions (see also  Ref.\cite{schonhammer1978time}) and reducing its numerical cost. 
However, to understand the advantages and limitations of the 
TD-SESCC and TD-DUCC formalisms in calculating physical observables, 
in the next section, we will analyze the properties of the corresponding action functionals. 
%}

\section{Action functionals for TD-SESCC and TD-DUCC formulations}
\label{section4}
% revision_kk
%{\color{red} Here goes the lithany about Lambda/T pairs!!! Mention problems of the TD-SESCC and advantages of TD-DUCC}

The form of optimal time evolution (or optimal equations of motion) for approximate theories describing 
time-dependent wavefunction $|\Psi\rangle$ can be determined by using the Dirac-Frenkel time-dependent variational 
principle (TDVP),\cite{frenkel1934wave,kramer1980geometry} (for a detailed discussion see also Refs.\cite{lowdin1972some,moccia1973time,
reinhard1977comment,kramer1980geometry,meyer1990multi,broeckhove1988equivalence,goings2018real}) 
where  one varies the action integrals
\begin{equation}
\mathcal{S}[\Psi(t)] = \int_{t_1}^{t_2} \langle\Psi(t)|i\hbar \frac{\partial}{\partial t} - H |\Psi(t)\rangle dt \;.
\label{dirfren}
\end{equation}
to obtain the equations of motion for parameters describing time-dependent wave function.
This functional 
has paved the wave for time-dependent  formulations, including various variants of 
time-dependent  multi-configurational  methods.\cite{meyer1990multi,miranda2011multiconfigurational,hochstuhl2012time,sato2013time,sato2015time,miyagi2014time,miyagi2014time2}  In the above formulation, the normalization of 
the time-dependent wave function,
\begin{equation}
\langle\Psi(t)|\Psi(t)\rangle = 1 \;,
\label{normal1}
\end{equation}
is assumed, which significantly simplifies the utilization of TDVP since for this type of domain the 
$i\hbar \frac{\partial}{\partial t}$ operator is the Hermitian operator and  $\mathcal{S}[\Psi(t)] $ functional assumes 
real values.
%time-dependent restricted-active-space configuration interaction method,\cite{rasci}
The  Dirac-Frenkel functional has been extended  to  bi-variational CC  formulations by Arponen \cite{arponen1983variational}
%% revision_kk
%{\color{blue} 
(for original discussion of bi-variational formalism see Ref.\cite{chernoff2006properties})
%}
\begin{equation}
\mathcal{S}[\Psi(t),\Psi'(t)]=
 \int_{t_1}^{t_2} \langle\Psi'(t)|i\hbar \frac{\partial}{\partial t} -H |\Psi(t)\rangle dt \;.
\label{arpo1}
\end{equation}
where the trial wave functions $\langle\Psi'(t)|$ and $|\Psi(t)\rangle$  satisfy the normalization condition 
\begin{equation}
\langle \Psi'(t)|\Psi(t)\rangle = 1 \;.
\label{inc1}
\end{equation}
%% revision_kk
%{\color{blue} 
The analysis of the CC methods is greatly facilitated by using the  Lagrangian  (see Ref.\cite{kramer1980geometry}) defined as 
\begin{equation}
L(\Psi(t),\Psi'(t))= \langle\Psi'(t)|i\hbar \frac{\partial}{\partial t} -H |\Psi(t)\rangle \;,
\label{lagr1}
\end{equation}
and 
\begin{equation}
\mathcal{S}[\Psi(t),\Psi'(t)]=
 \int_{t_1}^{t_2} 
 L(\Psi(t),\Psi'(t)) dt \;.
\label{lagr2}
\end{equation}
% revision2_kk
%{\color{blue}
The Lagrangian  (\ref{lagr1}), for simplicity denoted as $L(\Psi(t),\Psi'(t))$, should be understood in a broader context not only  as a function of $\langle\Psi'(t)|$
and $|\Psi(t)\rangle$ states but also as a function of their time-derivatives. 
%}
%
The Arponen's functional provides a  theoretical foundation, especially for time-dependent formulations of the
coupled cluster theory. Recently, Kvaal and  Pedersen used this functional to develop orbital adaptive time-dependent 
coupled cluster approximations \cite{kvaal2012ab}  and symplectic integrators for TD-CC equations.\cite{pedersen2019symplectic} 
The real-action functional 
introduced by Sato {\it et al.} in studies of TD-CC theory also invokes the bi-orthogonal form of the Arponen's 
functional.\cite{sato2018communication}
Both functionals (\ref{dirfren}) and (\ref{arpo1}) are very useful in situations where   one also  considers 
time evolution of reference function and molecular basis.

In this section, we will employ functionals (\ref{dirfren})  
and (\ref{arpo1}) to study stationary conditions for TD-SESCC and TD-DUCC 
formulations, respectively, and  to describe time evolution of sub-system described by the 
internal type excitations. 
The Lagrangian for normalized DUCC wave function  takes the following form 
\begin{eqnarray}
L[\sigma_{\rm int}(t),\sigma_{\rm ext}(t)] &=& \nonumber \\
&&\hspace*{-2.9cm} \langle\Phi| e^{-\sigma_{\rm int}(t)} e^{-\sigma_{\rm ext}(t)} 
(i\hbar \frac{\partial }{\partial t} - H) e^{\sigma_{\rm ext}(t)} e^{\sigma_{\rm int}(t)} |\Phi\rangle
\label{duccfd}
\end{eqnarray}
where, as stated earlier, we assume that one-particle orbital basis is time-independent. 
Since the DUCC wave function is normalized, the $S[\sigma_{\rm int}(t),\sigma_{\rm ext}(t)]$ assumes real values. 
In the next step, we will re-write functional (\ref{duccfd}) using results  of the previous section  
%(\ref{section3}) 
(see  Eqs. (\ref{sned2}), (\ref{bubu1}), (\ref{dera}), (\ref{asigt})) 
\begin{equation}
i\hbar e^{-\sigma_{\rm ext}(t)} \frac{\partial }{\partial t}  e^{\sigma_{\rm ext}(t)} = i\hbar \frac{\partial }{\partial t}
+i\hbar A(\sigma_{\rm ext}(t), \frac{\partial \sigma_{\rm ext}(t)}{\partial t})\;,
\label{bbtt}
\end{equation}
which after substituting to Eq.(\ref{duccfd}) leads to the expression for the Lagrangian
\begin{widetext}
\begin{equation}
\begin{aligned}
L[\sigma_{\rm int}(t),\sigma_{\rm ext}(t)]=\langle\Phi| e^{-\sigma_{\rm int}(t)} 
\lbrace i\hbar \frac{\partial }{\partial t} - [\bar{H}_{\rm ext}(t) -i\hbar A(\sigma_{\rm ext}(t), \frac{\partial \sigma_{\rm ext}(t)}{\partial t}) ] 
\rbrace e^{\sigma_{\rm int}(t)} |\Phi\rangle \;.
\label{duccfd2}
\end{aligned}
\end{equation}
%\end{widetext}
The above expression  can be represented in an equivalent form using  (\ref{gammae}) and (\ref{pqint})
%\begin{widetext}
\begin{equation}
\begin{aligned}
L[\sigma_{\rm int}(t),\sigma_{\rm ext}(t)]=\langle\Phi| e^{-\sigma_{\rm int}(t)} 
\lbrace i\hbar \frac{\partial }{\partial t} -  H^{\rm eff}(\sigma_{\rm ext}(t), \frac{\partial \sigma_{\rm ext}(t)}{\partial t},t) 
\rbrace e^{\sigma_{\rm int}(t)} |\Phi\rangle \;,
\label{duccfd3}
\end{aligned}
\end{equation}
\end{widetext}
which can be interpreted  in terms of slow- and fast-varying parts of the wave function represented in terms of 
$\sigma_{\rm int}(t)$ and $\sigma_{\rm ext}(t)$ operators if the "energetic" definition of the active space (or SES) is utilized.
Namely,
{\it if the fast-varying in time part of the wave function (or $\sigma_{\rm ext}(t)$-dependent part of the wave function) is known or 
can be efficiently approximated then the slow-varying  dynamic (captured by the proper  choice of the active space and $\sigma_{\rm int}(t)$ operator)
 of the entire system can be described as a sub-system dynamics generated by the Hermitian
$H^{\rm eff}(\sigma_{\rm ext}(t), \frac{\partial \sigma_{\rm ext}(t)}{\partial t})$ operator.}  This decoupling of various time regimes (slow- vs. fast-varying components)  
is analogous to the decoupling of high- and low-energy Fermionic degrees of freedom in stationary formulations of the SESCC and DUCC formalisms 
(see Refs.\cite{bauman2019downfolding,bauman2019quantum}). 
This result  indicates  that there exist  universal mechanisms in coupled cluster theory that naturally lead to the decoupling of  
various temporal, energy,  and spatial scales as shown  in present analysis and in 
Refs.\cite{safkk,bauman2019downfolding}. 

Properties  of the Arponen's functional draw heavily on the representations of the bra and ket states. Here we will analyze the standard approach 
analyzed in references \cite{arponen1983variational,kvaal2012ab,pigg2012time,sato2018communication,pedersen2019symplectic} 
where ket state is represented by standard CC expansion (\ref{ccexp}) while the bra is represented using a standard de-excitation $\Lambda(t)$ operator \cite{arponen1983variational}
\begin{eqnarray}
%\begin{aligned}
&&\langle\Psi'(t)| = \langle\Phi|(1+\Lambda(t)) e^{-T(t)}  
%\nonumber \\
\label{guu1} \\  
&&=  \langle\Phi|(1+\Lambda_{\rm int}(t)+\Lambda_{\rm ext}(t)) 
e^{-T_{\rm int}(t)} e^{-T_{\rm ext}(t)} \;, \label{guu2}
%\end{aligned}
\end{eqnarray}
which leads to the following form of the Lagrangian for the  Arponen's functional
\begin{widetext}
\begin{eqnarray}
L[\Lambda_{\rm int}(t),\Lambda_{\rm ext}(t),T_{\rm int}(t),T_{\rm ext}(t)] = \langle\Phi|(1+\Lambda_{\rm int}(t)+\Lambda_{\rm ext}(t)) 
e^{-T_{\rm int}(t)} e^{-T_{\rm ext}(t)}
(i\hbar \frac{\partial }{\partial t} -H) 
e^{T_{\rm ext}(t)} e^{T_{\rm int}(t)} |\Phi\rangle  &&\label{fqq1} \\
= \langle\Phi|(1+\Lambda_{\rm int}(t))e^{-T_{\rm int}(t)} (i\hbar \frac{\partial }{\partial t} -\bar{H}_{\rm ext}(t)) e^{T_{\rm int}(t)} |\Phi\rangle
+\langle\Phi|\Lambda_{\rm ext} (t)e^{-T_{\rm int}(t)}  (i\hbar \frac{\partial T_{\rm ext}(t)}{\partial t} + i\hbar \frac{\partial }{\partial t}
-\bar{H}_{\rm ext})e^{T_{\rm int}(t)} |\Phi\rangle && \;. \label{fqq2}
\end{eqnarray}
\end{widetext}
Comparing the above expression for the current functional with the DUCC one (\ref{duccfd3})
leads to the observation that the DUCC functional  can be expressed directly  in terms of a sub-system perspective, where the dynamics for the sub-system 
is generated by $\bar{H}_{\rm ext}-i\hbar A$ whereas for the TD-SESCC case only the first term in Eq.(\ref{fqq2}) can be interpreted 
in a similar way. The remaining term in Eq.(\ref{fqq2}) introduces explicit coupling between  $\Lambda_{\rm ext}(t)$ and
$T_{\rm int}(t)$ operators and cannot be considered in terms of sub-system dynamics, as in the case of DUCC functional.
% revision_kk
%{\color{blue}
This is a consequence of the fact that the 
state (\ref{guu1}) is not multiplicatively separable.
%}
As mentioned earlier, the algebraic form and properties of Arponen's functional depends on the parametrization used for the 
bra state $\langle\Psi'(t)|$. In the appendix, we analyze the form of the Arponen's functional where the extended coupled cluster (ECC) method 
\cite{arponen1983variational} is invoked to represent $\langle\Psi'(t)|$ state 
% reference_kk
%{\color{blue}
(see also Ref.\cite{kvaal2020guaranteed} for recent developments of the time-dependent ECC formalism). 
%} 
%%
We show that  it is possible to derive a similar form
of expressions as in the DUCC case.
%revision_kk
%however, the calculation of the corresponding downfolded  Hamiltonian poses a significant  challenge due to its complicated
many-body structure. \\
% these results hinge upon 

%
%
%
%  Decomposition of both X and T into int and ext parts.
%
%
%
%
%
%
% revision_kk - change of section title 
\section{Imaginary time evolution generated by downfolded Hamiltonians}
\label{section5}
%revisin_kk
%Before discussing main approximations of TD-SESCC and TD-DUCC approaches, 
%%% revision_kk
%{\color{blue} 
Due to the Hermitian character of the effective Hamiltonian   and  potential applications in quantum computing in this section we will focus entirely on the DUCC approach.
%}
In this section, we will  focus on the stationary conditions for the  $\sigma_{\rm int}(t)$ operator 
stemming from functional (\ref{duccfd2}). 
Since the  $e^{\sigma_{\rm int}(t)}|\Phi\rangle$ is an exact wave function in the complete active space, it can be equivalently 
represented by the active-space FCI expansion 
\begin{equation}
|\Psi_{\rm int}(t)\rangle = e^{\sigma_{\rm int}(t)}|\Phi\rangle = C_{\rm int}(t)|\Phi\rangle \;,
\label{uccci}
\end{equation}
where $C_{\rm int}(t)$ is the configuration interaction type operator that produces all possible excitations/configurations within CAS and 
which satisfies the normalization condition
\begin{equation}
\langle\Psi_{\rm int}(t)|\Psi_{\rm int}(t)\rangle = \langle\Phi|C_{\rm int}(t)^{\dagger} C_{\rm int} |\Phi\rangle = 1 \;.
\label{cici}
\end{equation}
In the
% $(\sigma_{\rm int}(t),sigma_{\rm ext}(t))$ parametrization of functional 
$(C_{\rm int}(t),\sigma_{\rm ext}(t))$ parametrization, the Lagrangian (\ref{duccfd2}) takes the form 
\begin{widetext}
\begin{equation}
L[C_{\rm int}(t),\sigma_{\rm ext}(t)]=\langle\Phi| C_{\rm int}(t)^{\dagger}
\lbrace i\hbar \frac{\partial }{\partial t} - [\bar{H}_{\rm ext}(t) -i\hbar A(\sigma_{\rm ext}(t), \frac{\partial \sigma_{\rm ext}(t)}{\partial t}) ] 
\rbrace C_{\rm int}(t) |\Phi\rangle \;.
\label{duccfd23}
\end{equation}
\end{widetext}
One should notice that there is no change in the algebraic form of the $\bar{H}_{\rm ext}(t)$ and 
$A(\sigma_{\rm ext}(t), \frac{\partial \sigma_{\rm ext}(t)}{\partial t})$ operators and stationary conditions for $C_{\rm int}(t)$ can be 
obtained following derivations for the TD-MCSCF and TD-CASSCF procedures of Refs.\cite{miranda2011multiconfigurational,sato2013time}, i.e.,
\begin{equation}
i\hbar \frac{\partial C_{\rm int}(t)}{\partial t}|\Phi\rangle = 
  H^{\rm eff} (\sigma_{\rm ext}(t), \frac{\partial \sigma_{\rm ext}(t)}{\partial t}) 
C_{\rm int}(t)|\Phi\rangle \;,
\label{ciintt}
\end{equation}
where the $H^{\rm eff}(\sigma_{\rm ext}(t), \frac{\partial \sigma_{\rm ext}(t)}{\partial t})$ operator is given by Eq.(\ref{gammae}). 
The above equations correspond to the 
equations (see Eq.(\ref{heffrev1}))  defining dynamics of the $\sigma_{\rm int}(t)$ operator when returning to the
$(\sigma_{\rm int}(t),\sigma_{\rm ext}(t))$ parametrization.
%% review_kk
%{\color{blue}
Since exponential parametrization (which assures proper normalization of the sub-system wave function) is replaced by its linear representation, in numerical algorithms to solve (\ref{ciintt}) one needs to re-normalized 
$C_{\rm int}(t)|\Phi\rangle$ in each step using techniques of 
Refs.\cite{kato2004time,sato2013time,sato2015time}.
%}
% review_kk
In view of decoupling internal and external wave function parameters in the $H^{\rm eff} (\sigma_{\rm ext}(t), \frac{\partial \sigma_{\rm ext}(t)}{\partial t})$ operator
a natural approximation is to use perturbative estimates of $\sigma_{\rm ext}(t)$ and describe evolution,
of $C_{\rm int}(t)|\Phi\rangle$ or $e^{\sigma_{\rm int}(t)}|\Phi\rangle$  (Eq.(\ref{ciintt}) or (\ref{heffrev1})) using  
modified  TD-MCSCF or  TD-CASSCF implementations.
\cite{meyer1990multi,miranda2011multiconfigurational,sato2013time,miyagi2014time}
In these simulations, the time-dependent $H^{\rm eff} (\sigma_{\rm ext}(t), \frac{\partial \sigma_{\rm ext}(t)}{\partial t})$ operator is used to generate the sub-system dynamics, which can be employed 
for example in studies of decoherence effects in open-systems. This can be achieved by analyzing density matrix corresponding 
to $C_{\rm int}|\Phi\rangle$ wave function. 

% revision_kk
%Important features of the TD-DUCC formalism are associated with the behavior of the above formalism during imaginary time evolutions of equations 
%(\ref{ciintt}) for $C_{\rm int}(t)$ amplitudes. 
%In analogy to the discussion by Pigg {\it et al.} \cite{pigg2012time} 
%
% revision2_kk
%{\color{blue} 
In order to discuss imaginary time evolution based on the utilization of effective Hamiltonians we introduce  imaginary 
Wick rotation $t \rightarrow -i\hbar \tau$. However, to take care of proper normalization of the $C_{\rm int}(\tau)|\Phi\rangle$ expansion in the $\tau \rightarrow \infty$  limit,
one has to introduce appropriate renormalization term (or the so-called energy shift). 
Similar procedures are used in the context of other methods based on  the imaginary-time Schr\"odinger equation, for example stochastic FCI methods, \cite{booth2009fermion} and recent formulations of quantum  imaginary time evolution.\cite{mcardle2019variational}
To define the equations for $C_{\rm int}(\tau)$ one can consider a steepest-descent flow for energy expectation value for normalized $C_{\rm int}(\tau)|\Phi\rangle$ function and effective Hamiltonian. This procedure leads to the following equations:
\begin{widetext}
\begin{equation}
 \frac{\partial C_{\rm int}(\tau)}{\partial \tau}|\Phi\rangle = 
-  ( H^{\rm eff} (\sigma_{\rm ext}(\tau), \frac{\partial \sigma_{\rm ext}(\tau)}{\partial \tau}) - S(\tau)) 
C_{\rm int}(\tau)|\Phi\rangle \;,
\label{ciintt2}
\end{equation}
\end{widetext}
where the energy shift $S(\tau)$ is defined as a value of the  energy functional constructed for  $H^{\rm eff} (\sigma_{\rm ext}(\tau), \frac{\partial \sigma_{\rm ext}(\tau)}{\partial \tau})$ and properly normalized 
$C_{\rm int}(\tau)|\Phi\rangle$ trial function.
%} % revision2_kk
Assuming that  this approximation reaches the stationary limit (SL) for $\tau\rightarrow \infty$ (i.e.  
$\frac{\partial C_{\rm int}(\tau)}{\partial \tau}=0$,  $\frac{\partial \sigma_{\rm ext}(\tau)}{\partial \tau}=0$)
and 
\begin{eqnarray}
\sigma_{\rm ext} (\tau) &\xrightarrow{SL}& \sigma_{\rm ext} \;, \label{statio2}
\end{eqnarray}
then the form of Eq.(\ref{ciintt2}) for large values of $\tau$ is given by a simpler form 
%
% revision2_kk
%{\color{blue} 
\begin{widetext}
\begin{equation}
 \frac{\partial C_{\rm int}(\tau)}{\partial \tau}|\Phi\rangle = 
-  (P+Q_{\rm int}) (\bar{H}_{\rm ext} (\sigma_{\rm ext}(\tau)))- \bar{S}(\tau)) (P+Q_{\rm int})  C_{\rm int}(\tau)|\Phi\rangle \;,
\label{ciintt3}
\end{equation}
\end{widetext}
where $\bar{S}(\tau)$ is a value of energy functional defined by the 
$(P+Q_{\rm int}) (\bar{H}_{\rm ext} (\sigma_{\rm ext}(\tau)) (P+Q_{\rm int})$ operator  and  normalized 
$C_{\rm int}(\tau)|\Phi\rangle$ trial function.
%}  
% revision2_kk
This step  is a consequence of the fact that the "external-velocity" dependent term (i.e. $\frac{\partial \sigma_{\rm ext}(\tau)}{\partial \tau}$-dependent term  - the 
$A$ operator in Eq.(\ref{asigt})) disappears in the SL limit, i.e.,
\begin{eqnarray}
A(\sigma_{\rm ext}(\tau), \frac{\partial \sigma_{\rm ext}(\tau)}{\partial \tau})  &\xrightarrow{SL}& 0 \;. \label{statio3}
\end{eqnarray}
This important result is intuitively in agreement with the DUCC analysis discussed in previous papers \cite{bauman2019downfolding,bauman2019quantum,bauman2020coupled} where it was shown that the energy of the system can be obtained by diagonalizing stationary $\bar{H}_{\rm ext}$  Hamiltonian in the active space. 
Additionally, $\bar{H}_{\rm ext}(\sigma_{\rm ext}(\tau))$ is a similarity transformed Hamiltonian $H$ which is bounded from below.
Projecting this Hamiltonian on a $(P+Q_{\rm int})$ subspace produces a downfolded Hamiltonian, which is bounded from below. 
This analysis shows the feasibility of imaginary time evolution in the TD-DUCC case. 
Additionally, since in the SL $\bar{H}_{\rm ext}(\sigma_{\rm ext}(\tau))$ approaches $\bar{H}_{\rm ext}$, to get the ground state energy 
one can alternatively perform imaginary evolutions with the fixed-in-time Hamiltonian $\bar{H}_{\rm ext}$ in the active space.

%1.) fixing T(ext) and doing Gunnarson, Hoodbhoy,  Pigg and Kuhl algorithms  2.) MCSCF for DUCC
%propagation of $T_{\rm int}(t)$ in fixed or predefined $T_{\rm ext}(t)$ frame (AS IN PAPENBROCK'S PAPER). Possibility of "communicating" two various active-spaces defined by sum of two functionals. CHECK IT WHAT IMPACT IT HAS. OPEN SYSTEMS - for the subsystem (an non-optimized w.r.t. external amplitudes energy is not conserved) . Open system description .
%
%time-dependence of Gamma - is energy conserved once we optimize only 
%Gamma - bounded operator --> sub-space representation of $\hbar{H}_{ext}$
%$\hbar{H}_{ext}$ the same eigenvalues as H for all t 

\section{Conclusions} 
In this paper, we analyzed the extension of time-independent SESCC and DUCC formalisms to the time domain. 
% revision_kk
%{\color{blue}
We also proved the exactness of the double unitary CC Ansatz based on the general-type anti-Hermitian internal and external cluster operators.
%}
%
For a fixed-in-time orbital basis, we were able to prove that it is possible to describe the dynamics of the entire system for the ket states in 
terms of sub-system dynamics generated by a downfolded Hamiltonian. 
%revision_kk
%{\color{blue} 
This feature, in the case of the SESCC formulation provides a way for reducing the numerical cost of CC Green's function formulations.
%}
%
%revision_kk
%{\color{blue}
For the DUCC and ECC approaches we also showed the dynamic separability at the level of action functional, which is especially important from the point of view of constructing approximate formalisms.
%}
If the active space  separates 
slow modes from the  fast ones, perturbative techniques can be used to define downfolded Hamiltonians.  
%Although these results are valid for both TD-SESCC and TD-DUCC methods,  
In this paper, we put special emphasis on the TD-DUCC formalism 
where the operators defining the sub-system dynamics are Hermitian.
Recognizing that if the $\sigma_{\rm ext}(\tau)$ operator reaches the stationary limit, then the time-dependent 
Hamiltonian converges to the time-independent downfolded Hamiltonian, demonstrating the feasibility of imaginary time evolution. In the stationary limit, the $A$ operator, which 
depends linearly on the 
$\sigma_{\rm ext}(t)$-velocity  term,  disappears, enabling the utilization of the 
time-dependent  CASSCF codes to describe system dynamics in the time domain  captured 
by the specific choice of the active space. Additionally,  these results allow quantum Lanczos algorithms 
to be employed to identify the ground-state  energy values.\cite{motta2020determining}
It is also interesting to look at the sub-system dynamics from the point of view of description decoherence effects, where the sub-system of interest 
is in contact with the surrounding environment described by the $\bar{H}^{\rm eff}(t)$ operator. 
% revision_kk
%{\color{blue}
In this paper, we cannot cover all aspects of the dynamics separation. Therefore in the following papers, we will study related topics such as time evolution of extensive observables, ECC formalisms, and DUCC approximate approaches for classical and quantum computing.
%} 
%%
% 
%

\section{Data Availability Statement}

The data that support the findings of this study are available from the corresponding author upon reasonable request.

\section{Acknowledgement}
This  work  was  supported  by  the "Embedding Quantum Computing into Many-body Frameworks for Strongly Correlated  Molecular and Materials Systems" project, 
which is funded by the U.S. Department of Energy(DOE), Office of Science, Office of Basic Energy Sciences, the Division of Chemical Sciences, Geosciences, and Biosciences.
All calculations have been performed using computational resources  at the Pacific Northwest National Laboratory (PNNL).
PNNL is operated for the U.S. Department of Energy by the Battelle Memorial Institute under Contract DE-AC06-76RLO-1830.

%\appendix
%\begin{subappendices}
\begin{appendices}
\section{Sub-system dynamics with extended coupled cluster formalism}

As a specific example of the SESCC Arponen's functional,  we will focus on  the bi-orthogonal formulation of TDVP 
given by the  extended coupled cluster formalism 
where bra and ket trial states $\langle\Psi'(t)|$ and   $|\Psi(t)\rangle$  are parametrized as
\begin{eqnarray} 
\langle\Psi'(t)| &=& \langle\Phi| e^{X(t)}  e^{-T(t)} \;,  \label{ecc1}  \\
|\Psi(t) \rangle &=& e^{T(t)} |\Phi\rangle \;, \label{ecc2} 
\end{eqnarray}
where $T(t)$ and $X(t)$ are standard excitation and de-excitation operators. 
Expressions  (\ref{ecc1}) and (\ref{ecc2})  for trial wave function  automatically satisfies 
normalization condition (\ref{inc1})
and the  
functional (\ref{arpo1}) can be viewed as a function of cluster amplitudes defining 
$T(t)$ and $X(t)$ cluster operators
\begin{equation}
\mathcal{S}[T(t),X(t)]=
 \int_{t_1}^{t_2} \langle\Phi|e^{X(t)} e^{-T(t)} (i\hbar \frac{\partial}{\partial t} - H) e^{T(t)} |\Phi\rangle dt \;.
\label{arpo12}
\end{equation}
The stationary conditions provide equations of motion for the cluster amplitudes $T(t)$ and $X(t)$.
Moreover, partial integration of (\ref{arpo12}) is required to obtain equations of motion for the $X(t)$ operator.

\begin{figure}
\includegraphics[width=0.45 \textwidth]{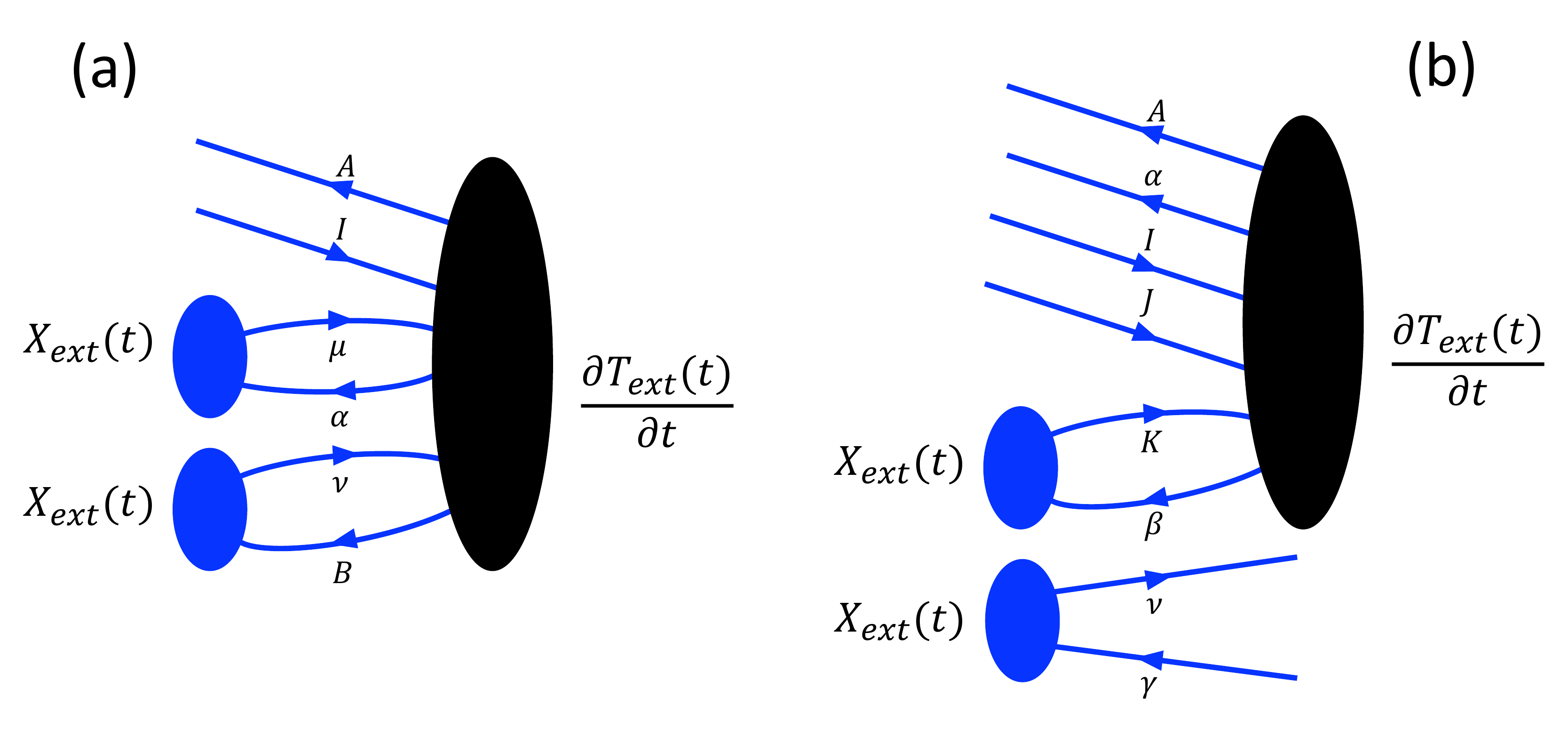}
\caption{ Examples of connected (a) and disconnected (b) diagrams contributing to $e^{X_{\rm ext}(t)} \frac{\partial T_{\rm ext}(t)}{\partial t}$. While the  former diagram may result in a non-zero contributions to the first term on the right-hand side of (\ref{idt2}) the latter one gives zero contribution.
}
\label{fig2}
\end{figure}

In analogy to standard excitation operator $T$, 
we will partition the de-excitation operator into its internal and external components (now the notion of sub-system embedding excitations  should be replaced by the
sub-system embedding de-excitations) 
\begin{equation}
X(t)=X_{\rm int}(t)+X_{\rm ext}(t)
\end{equation}
where all strings of de-excitation operators defining $X_{\rm int}(t)$ contain active spin-orbital indices only. 
The Arponen's functional for the ECC formalism takes the form 
\begin{widetext}
\begin{eqnarray}
\mathcal{S}[T_{\rm int}(t),T_{\rm ext}(t),X_{\rm int}(t),X_{\rm ext}(t)] = && \nonumber \\
 \int_{t_1}^{t_2} \langle\Phi|e^{X_{\rm int}(t)} e^{X_{\rm ext}(t)} e^{-T_{\rm int}(t)} e^{-T_{\rm ext}(t)}(i\hbar \frac{\partial}{\partial t} - H) 
 e^{T_{\rm ext}(t)}  e^{T_{\rm int}(t)} |\Phi\rangle dt  \;, 
 \label{aecc}
\end{eqnarray}
\end{widetext}
which is composed of time derivative $\mathcal{S}_{Dt}[T_{\rm int}(t),T_{\rm ext}(t),X_{\rm int}(t),X_{\rm ext}(t)]$ part and 
energy term $\mathcal{S}_H[T_{\rm int}(t),T_{\rm ext}(t),X_{\rm int}(t),X_{\rm ext}(t)]$ ($\mathcal{S}=\mathcal{S}_{Dt}+\mathcal{S}_H$).
%In contrast to the real-valued DUCC functional, the expression above poses a significantly bigger challenge in extracting 
%sub-system dynamics as shown in Eq.(\ref{duccfd3}). 
%
%There are two main reasons 
%for that: (1) there are groups of non-commuting exponential operators, and (2) in the general case, functional (\ref{aecc}) can assume complex values.
%While the latter problem has been recently explored by introducing real-action functional \cite{sato2018communication} 
%(see also Ref.\cite{pedersen2019symplectic}), in order to define sub-system dynamic we will focus entirely on the first issue. 
%% revision_kk
%{\color{blue} 
To define sub-system dynamics 
%}
let us re-write $\mathcal{S}_{Dt}$ and  $\mathcal{S}_{H}$ energy functional  terms of (\ref{aecc}) with  Lagrangians
($L_{Dt}$ and $L_H$, $\mathcal{S}= \int_{t_1}^{t_2} (L_{Dt}(t)+L_H(t)) dt$). 
For $L_{Dt}(t)$ we have 
\begin{widetext}
\begin{eqnarray}
\hspace*{-0.4cm}
L_{Dt}(t) &=& \langle\Phi|e^{X_{\rm int}(t)}  e^{X_{\rm ext}(t)} e^{-T_{\rm int}(t)} e^{-T_{\rm ext}(t)} (i\hbar \frac{\partial}{\partial t} )
e^{T_{\rm ext}(t)}  e^{T_{\rm int}(t)} |\Phi\rangle \label{idt1} \\
&=& i\hbar \langle\Phi|e^{X_{\rm int}(t)}  e^{X_{\rm ext}(t)} \frac{\partial T_{\rm ext}(t)}{\partial t} |\Phi\rangle +
        \langle\Phi|e^{X_{\rm int}(t)} e^{-T_{\rm int}(t)} (i\hbar \frac{\partial}{\partial t}) e^{T_{\rm int}(t)} |\Phi\rangle \label{idt2}    \\
&=& i\hbar \langle\Phi|e^{X_{\rm int}(t)}  e^{-T_{\rm int}(t)} e^{T_{\rm int}(t)} (e^{X_{\rm ext}(t)} \frac{\partial T_{\rm ext}(t)}{\partial t})_C e^{-T_{\rm int}(t)} e^{T_{\rm int}(t)} |\Phi\rangle +
        \langle\Phi|e^{X_{\rm int}(t)} e^{-T_{\rm int}(t)} (i\hbar \frac{\partial}{\partial t}) e^{T_{\rm int}(t)} |\Phi\rangle \label{idt3}    \\        
&=& i\hbar \langle\Phi|e^{X_{\rm int}(t)}  e^{-T_{\rm int}(t)} B(T_{\rm int}(t),\frac{\partial T_{\rm ext}(t)}{\partial t},X_{\rm ext}(t))  e^{T_{\rm int}(t)} |\Phi\rangle +
        \langle\Phi|e^{X_{\rm int}(t)} e^{-T_{\rm int}(t)} (i\hbar \frac{\partial}{\partial t}) e^{T_{\rm int}(t)} |\Phi\rangle \label{idt4} 
\end{eqnarray}
\end{widetext}
where $B(T_{\rm int}(t),\frac{\partial T_{\rm ext}(t)}{\partial t},X_{\rm ext}(t))$ (or $B$ term for short)  is given by the formula
\begin{equation}
B=e^{T_{\rm int}(t)} (e^{X_{\rm ext}(t)} \frac{\partial T_{\rm ext}(t)}{\partial t})_C e^{-T_{\rm int}(t)} \;.
\label{bterm}
\end{equation}
% revision_kk 
%{\color{blue}  
The connected character of the $e^{X_{\rm ext}(t)} \frac{\partial T_{\rm ext}(t)}{\partial t}$ expression in the first term on the right-hand side of Eq.(\ref{idt3}) stems from the fact that in Eq.(\ref{idt2})  the $e^{X_{\rm ext}(t)} \frac{\partial T_{\rm ext}(t)}{\partial t}$ operator is bracketed by two states belonging 
to the active space, i.e., $\langle\Phi|e^{X_{\rm int}(t)}$ and $|\Phi\rangle$, therefore for $e^{X_{\rm ext}(t)} \frac{\partial T_{\rm ext}(t)}{\partial t}$ to produce non-zero contributions all inactive lines have to be contracted out 
and all external lines of the resulting diagrams have to be active (see Fig.\ref{fig2}).
Consequently, only the active-space part of the $B$ operator produces a non-zero contribution to Eq.(\ref{idt4}).
%}
% revision_kk
%Notice that only the active-space part of the $B$ operator produces a non-zero contribution to Eq.(\ref{idt4}), and therefore 
%only the connected part of $e^{X_{\rm ext}(t)} \frac{\partial T_{\rm ext}(t)}{\partial t}$ produces   non-zero contributions to the first term on the right hand side of Eq.(\ref{idt3})
%(disconnected components contain uncontracted external  spin-orbital label which gives zero contribution when acting on either the active-space vector 
%$\langle\Phi|e^{X_{\rm int}(t)}$ or reference function $|\Phi\rangle$).

Using similar manipulations, the $L_{H}(t)$ term can be re-written as 
\begin{widetext}
\begin{eqnarray}
L_{H}(t) &=& 
 \langle\Phi|e^{X_{\rm int}(t)}  e^{X_{\rm ext}(t)} e^{-T_{\rm int}(t)} e^{-T_{\rm ext}(t)} 
H e^{T_{\rm ext}(t)}  e^{T_{\rm int}(t)} |\Phi\rangle \label{rrp1} \\
&=&  \langle\Phi|e^{X_{\rm int}(t)} e^{-T_{\rm int}(t)} \bar{H}_{ECC} e^{T_{\rm int}(t)} |\Phi\rangle \label{rrp2}
\end{eqnarray}
\end{widetext}
where
\begin{equation}
\bar{H}_{\rm ECC} =  e^{X^{\rm int}_{\rm ext}(t)} e^{-T_{\rm ext}(t)} H e^{T_{\rm ext}(t)} e^{-X^{\rm int}_{\rm ext}(t)} 
\label{rrp3}
\end{equation}
and 
\begin{equation}
X^{\rm int}_{\rm ext}(t)=e^{T_{\rm int}(t)} X_{\rm ext}(t) e^{-T_{\rm int}(t)}
\label{rrp4}
\end{equation}
is a similarity transformed $X_{\rm ext}(t)$ operator. 
%{\color{blue}
One should also notice that the above definition for the   $X^{\rm int}_{\rm ext}(t)$ operator assures its connected character.
%}

Summarizing, the $\mathcal{S}= \int_{t_1}^{t_2} (L_{Dt}(t)+L_H(t)) dt$ functional can be expressed as:
\begin{widetext}
\begin{equation}
 \mathcal{S}=\int_{t_1}^{t_2} \langle\Phi|e^{X_{\rm int}(t)} e^{-T_{\rm int}(t)} 
 \lbrace 
 i\hbar \frac{\partial}{\partial t} - [ \bar{H}_{\rm ECC}(t) - i\hbar B(t)]
 \rbrace
 e^{T_{\rm int}(t)} |\Phi\rangle \;.
\label{ubu}
\end{equation}
\end{widetext}
Although the  above formula bears a resemblance to  functional (\ref{duccfd2}), 
by their definitions  $\bar{H}_{\rm ECC}(t)$ and $B(t)$ mix internal and external Fermionic degrees 
of freedom, in contrast to the DUCC functional (\ref{duccfd2}). Another important feature of the ECC formalism is the fact that 
$\bar{H}_{\rm ECC}(t)$, as given by Eq.(\ref{rrp3}), is defined by rather complicated algebraic many-body structure defined by  two similarity transformations 
one involving singly-transformed auxiliary operator $X^{\rm int}_{\rm ext}(t)$. Efficient coding of these expressions may require sophisticated 
symbolic algebra tools. 
% revision_kk
%{\color{blue} 
This effort can be further facilitated by the utilization of the so-called doubly linked structure of resulting diagrams (see Ref.\cite{Arponen1987})
%}
%\end{subappendices}
\end{appendices}

%\bibliography{gfcc.bib}
%
%\end{document}


\end{document}